\documentclass[aps,prl,twocolumn,footinbib,amsmath,amssymb,amsfonts,superscriptaddress]{revtex4-2}

\usepackage{MnSymbol}
\usepackage{color,framed,graphicx,lipsum}
\usepackage{float}
\floatstyle{plaintop}
\restylefloat{table}
\usepackage{tikz}
\usepackage{tabularx}
\usepackage{upgreek}
\usepackage{dsfont}

\renewcommand{\thetable}{\arabic{table}}
\renewcommand{\figurename}{{Fig.}}
\usepackage{upgreek}

\allowdisplaybreaks

\newcommand{\DHK}[1]{\textcolor{black}{{#1}}}

\begin{document}
\title{Hierarchical entanglement shells of multichannel Kondo clouds}

\author{Jeongmin Shim}
\affiliation{Department of Physics, Korea Advanced Institute of Science and Technology, Daejeon 34141, Korea}
\affiliation{Present address: Arnold Sommerfeld Center for Theoretical Physics, Center for NanoScience, and Munich Center for Quantum Science and Technology, Ludwig-Maximilians-Universit\"at M\"unchen, 80333 Munich, Germany}
\affiliation{These authors contributed equally: Jeongmin Shim, Donghoon Kim}
 
\author{Donghoon Kim}
\affiliation{Department of Physics, Korea Advanced Institute of Science and Technology, Daejeon 34141, Korea}
\affiliation{These authors contributed equally: Jeongmin Shim, Donghoon Kim}

\author{H.-S. Sim}\email[e-mail: ]{hs_sim@kaist.ac.kr}
\affiliation{Department of Physics, Korea Advanced Institute of Science and Technology, Daejeon 34141, Korea}

\date{\today}

\begin{abstract}
Impurities or boundaries often impose nontrivial boundary conditions on a gapless bulk, resulting in distinct boundary universality classes for a given bulk, phase transitions, and non-Fermi liquids in diverse systems. The underlying boundary states however remain largely unexplored. 
This is related with a fundamental issue how a Kondo cloud spatially forms to screen a magnetic impurity in a metal.
Here we predict the quantum-coherent spatial and energy structure of multichannel Kondo clouds, representative boundary states involving competing non-Fermi liquids, by studying quantum entanglement between the impurity and the channels. Entanglement shells of distinct non-Fermi liquids coexist in the structure, depending on the channels. As temperature increases, the shells become suppressed one by one from the outside, and the remaining outermost shell determines the thermal phase of each channel. Detection of the entanglement shells is experimentally feasible. Our findings suggest a guide to studying other boundary states and boundary-bulk entanglement.
\end{abstract}
\maketitle
  
\begin{center}
\textbf{Introduction}
\end{center}

Boundary quantum critical phenomena~\cite{Vojta06,Diehl97} appear in gapless systems of quantum impurities~\cite{Vojta06,Nozieres78a,Nozieres78b,Gruner1974,Nozieres80,Andrei84,Affleck91b,Affleck93,Ludwig94,Hewson97}, magnets with surfaces~\cite{Laflorencie06}, edge states of topological orders~\cite{Fendley06}, and qubit dissipation~\cite{Leggett87,Cottet19}.  
There, the presence of a boundary causes various boundary criticalities that affect the bulk, depending on boundary-bulk coupling. A character of boundaries has been revealed by the boundary or impurity entropy~\cite{Affleck09a,Eriksson11,Cornfeld17,Alkurtass16} that is the entropy difference between the presence and absence of the boundary. This entropy corresponds to the constant term in the dependence of the ground-state entanglement entropy on the location of the entanglement partition~\cite{Alkurtass16}.
The entropy is a bulk quantity, as the partition is placed at long distance from the boundary,
and it has been obtained by using the boundary conformal field theory (BCFT)~\cite{Affleck91b,Affleck93,Ludwig94,Cardy06,Cardy89,Cardy91}, a standard approach for the criticalities.
 
While bulk quantities have been understood, boundary states are yet to be explored~\cite{Yoshida66,Affleck2010,Affleck14,Moca21}. The Kondo singlet~\cite{Yoshida66} in the single-channel Kondo effect, a many-body state of metallic electrons formed to screen a local impurity spin, implies that quantum entanglement between a bulk and its boundary is essential for understanding the quantum coherent boundary-bulk coupling~\cite{Lee15,Shim18,Yoo18}.
The spatial distribution of the particles forming the boundary-bulk entanglement will be a key information of boundary quantum criticalities and related many-body effects. 
As the partition for the boundary-bulk entanglement is placed right at the boundary~\cite{Lee15,Shim18,Yoo18,Kim21}, the entanglement differs from the boundary entropy.
There are difficulties in studying the entanglement.
In BCFTs, the boundary degrees of freedom are absorbed into the bulk as boundary conditions, and
bulk properties at long distance from the boundary are considered.
Experimentally detecting entanglement typically requires inaccessible multiparticle observables. Understanding about the entanglement is desired.

Multichannel Kondo effects,
where multiple channels of conduction electrons compete to screen an impurity spin,
serve as a paradigm of many-body physics and boundary criticalities~\cite{Nozieres80,Andrei84,Affleck91b,Affleck93,Ludwig94}.
For example, in the $k$-channel Kondo ($k$CK) effect, $k$ electron channels compete to screen an impurity spin 1/2. It is described by the Hamiltonian
\begin{align}
H_{k\mathrm{CK}} &= \sum_{j=1}^k  J_j \mathbf{S}_\mathrm{imp} \cdot \mathbf{S}_j(0) + \sum_{j=1}^k H_j. \label{MCK_bare}
\end{align}
Here, the impurity spin $\mathbf{S}_\mathrm{imp}$
locally couples to the spin $\mathbf{S}_j(0)$ of electrons in the $j$th channel with strength $J_j > 0$, and 
$H_{j}$ describes free electrons in the $j$th channel.
In the Affleck-Ludwig BCFT~\cite{Affleck91b,Affleck93,Ludwig94}, 
the channel isotropic case of $J_1 = \cdots = J_k$ is transformed into 
a free electron Hamiltonian with a nontrivial boundary condition,  by mapping $H_{j}$ to a semi-infinite one dimension,
and fusing the impurity with the boundary of the one dimension. It exhibits a boundary criticality.
In channel anisotropic cases,  the competition between the channels results in quantum phase transitions~\cite{Vojta06}, various non-Fermi liquids (NFLs)~\cite{Nozieres80,Affleck91b}, and fractionalizations~\cite{Affleck91a}, making the effects rich. 
Thermal phases and their renormalization flows of the channel anisotropic Kondo effects 
were experimentally observed by using quantum dots or metallic islands~\cite{Potok07,Keller15,Iftikhar15,Iftikhar18,DGG21}.  

The boundary states of the Kondo effects involve a Kondo cloud~\cite{Affleck2010,Affleck14,Moca21}
formed by the conduction electrons screening the impurity spin.
Theoretically the cloud has been studied~\cite{Eriksson11,Alkurtass16,Cornfeld17,Alternate_Proposal_1,Alternate_Proposal_2,Barzykin98} mostly for channel isotropic cases.
\DHK{For anisotropic 2CK effects, a quantity called the excess charge density was used to study a real-space structure that indicates spatial regions corresponding to the local moment and strong coupling phases~\cite{Mitchell11}. However this quantity can hardly quantify the spatial distribution of a Kondo cloud, as it can be negative at certain distances from the impurity spin and even increase with the distance.}  
The properties of the cloud, such as its channel-resolved spatial distribution, its entanglement with the impurity, its correspondence to the transition or crossover between distinct NFL phases, and its thermal suppression, are yet to be studied.
It also remains unknown how to detect the clouds in the multichannel cases,
while a cloud was recently observed~\cite{Park13,Borzenets20} in the single channel case.


The entanglement between an impurity and its Kondo cloud is a boundary-bulk entanglement~\cite{Lee15,Shim18,Yoo18,Kim21}. The spatial distribution of the electrons forming this entanglement will characterize how the cloud spatially screens the impurity quantum coherently. In this work, we propose how to theoretically quantify and experimentally measure the distribution by applying a perturbation of local symmetry breaking (LSB) at a distance from the impurity. The distribution is found to exhibit channel-dependent hierarchical entanglement shells of NFL, Kondo Fermi liquid (FL), or non-Kondo FL characters in the channel anisotropic cases.
Each shell is identified by a power-law decay of the distribution with the distance, whose exponent is determined by the scaling dimension of the boundary operator describing the character. As the temperature increases,
the shells are suppressed one by one from the outside,
and the remaining outmost shell determines the thermal phase of each channel.
The entanglement shell structure shows that different NFLs and FLs hierarchically coexist around the boundary with spatial and energetical separation, reflecting the renormalization of the quantum coherent impurity screening (quantified by the entanglement) in the presence of the channel competition.

\begin{figure}[t]
\centerline{\includegraphics[width=.9\columnwidth]{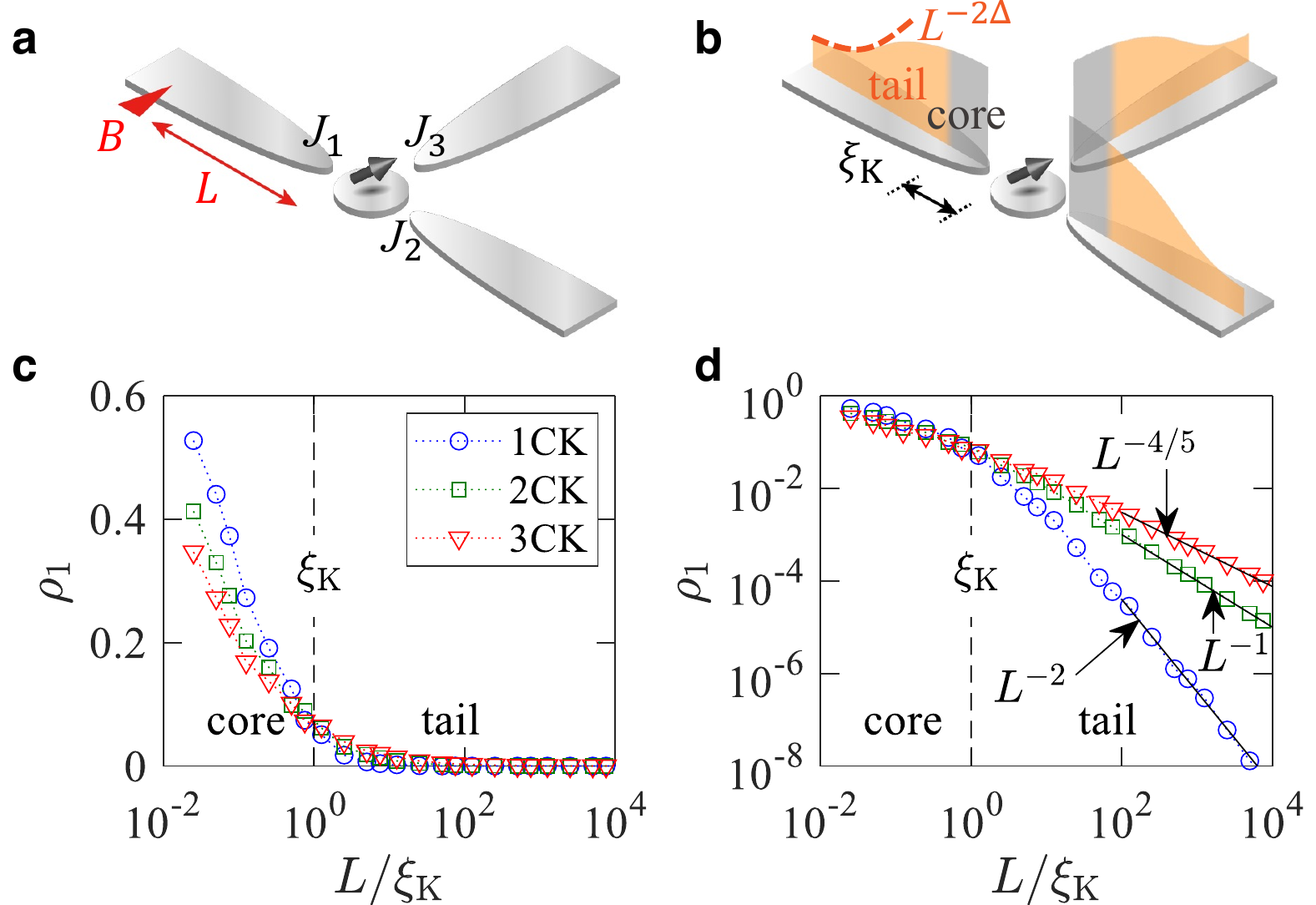}}
\caption{
{\bf Channel-isotropic Kondo cloud.} {\bf a} An impurity spin couples to three channels with equal strengths $J_1=J_2=J_3$. A perturbation $B$ breaks the SU(2) spin symmetry at distance $L$ from the impurity in channel 1. The cloud distribution $\rho_1 (L)$ in channel 1 is read out from the $L$ dependence of the entanglement  $\mathcal{N}$ between the impurity and the channels.
{\bf b} Schematic cloud distribution.  Crossover between the core and the tail happens around the cloud length $\xi_\textrm{K}$.
{\bf c} Numerical renormalization group (NRG) results of $\rho_1 (L)$ at zero temperature \DHK{for the isotropic single-channel (1CK), two-channel (2CK), and three-channel Kondo (3CK) effects.}
{\bf d} Log-log plot of {\bf c}. The tail follows the power-law decay $L^{-2 \Delta}$ in agreement with \DHK{the} boundary conformal field theory (BCFT).
}
\label{MCKC_fig1}
\end{figure}

\begin{center}
\textbf{Results}
\end{center}

\textbf{Quantifying boundary entanglement distribution ---} 
We study the entanglement negativity $\mathcal{N} \equiv\Vert \rho^{\mathrm{T}_{\mathrm{I}}} \Vert_1 - 1$ between the impurity and the channels in the $k$CK effects.
$\rho$ is the density matrix of the whole system, $\Vert \cdot \Vert_1$ is the trace norm, and $\mathrm{T}_{\mathrm{I}}$ means the partial transpose on the impurity. 
This negativity is  twice the conventional definition~\cite{Horodecki09,Vidal02} so that its maximum value is 1.
It measures quantum coherence of the screening.
The screening happens by
the maximal entanglement $\mathcal{N} = 1$ independent of $k$ in the channel isotropic cases at zero temperature~\cite{Kim21}. 


To quantify the spatial distribution of the entanglement, we apply an LSB perturbation breaking the Kondo SU(2) symmetry in a channel $n$ at distance $L$ from the impurity [Fig.~\ref{MCKC_fig1}a], and study the reduction $\rho_n$ of the negativity
from the value $\mathcal{N}_0 (T)$ in the absence of the LSB to $\mathcal{N} (L,T; n)$ in the presence of the LSB,
\begin{align}\label{eq:profile}
\rho_{n} (L,T) \equiv \mathcal{N}_0 (T) - \mathcal{N} (L,T; n),
\end{align}
at temperature $T$. $\rho_n$ varies between 0 and 1. Larger $\rho_n$ implies that at the distance $L$ there exist more electrons participating in the entanglement.
Therefore the $L$ dependence of the reduction $\rho_{n} (L,T)$ quantifies the spatial distribution of the Kondo cloud in the channel $n$.  

The negativity has a direct relation~\cite{Kim21} with the impurity magnetization 
$\mathbf{M} = \langle \mathbf{S}_\text{imp} \rangle$ at zero temperature (Supplementary Note 1),
\begin{equation}
\mathcal{N} = \sqrt{1 -  \frac{4 \mathbf{M}^2}{\hbar^2}}, \label{magnetization_zero}
\end{equation}
where $\mathbf{S}_\text{imp}$ is the impurity spin operator.
This shows that the magnetization is larger as the impurity spin is less screened by,
equivalently less entangled with, conduction electrons.
This relation is valid at zero temperature in general situations of the Kondo effects,
and it is a good approximation at low temperature $T \ll T_\textrm{K}$, where $T_\textrm{K}$ is the Kondo temperature.
    
 
For details, we consider a Hamiltonian $H_{k\mathrm{CK}}+H_\mathrm{LSB}$.
The Kondo Hamiltonian $H_{k\mathrm{CK}}$ is shown in Eq.~\eqref{MCK_bare}. 
	Here each channel is described by free electrons in a semi-infinite one dimensional system and the impurity spin is located at the boundary of the one dimension.
$H_\mathrm{LSB}$ describes the LSB by a local magnetic field $B$ along $x$ axis 
coupled to the spin $S_{n,x} (L)$  in a channel $n$ at distance $L$
from the impurity,
\begin{align}\label{eq:LSB}
H_\mathrm{LSB} &= B S_{n, x}(L).
\end{align}
In the presence of the LSB, we compute the negativity between the impurity and the channels at finite temperature by using the numerical renormalization group (NRG) method (Supplementary Notes 2-4) that we have developed~\cite{Shim18}. 
We also obtain the negativity at zero temperature by using Eq.~\eqref{magnetization_zero} and analytically computing  the magnetization based on the BCFT in the presence of the LSB (Supplementary Note 5).

\textbf{Isotropic multichannel Kondo clouds ---} 
We first consider the channel isotropic case of $J_1=J_2=\cdots=J_k=J$.
At $T \sim T_\mathrm{K}$, there occurs thermal crossover from the infrared Kondo fixed point  to the ultraviolet local moment (LM) phase.
The Kondo phase is a FL in the single-channel case~\cite{Nozieres78a,Nozieres78b} and a NFL in the multichannel cases of $k \ge 2$~\cite{Nozieres80,Affleck91b}.

\onecolumngrid
\begin{center}
\begin{figure*}[t]
\centerline{\includegraphics[width=.9\textwidth]{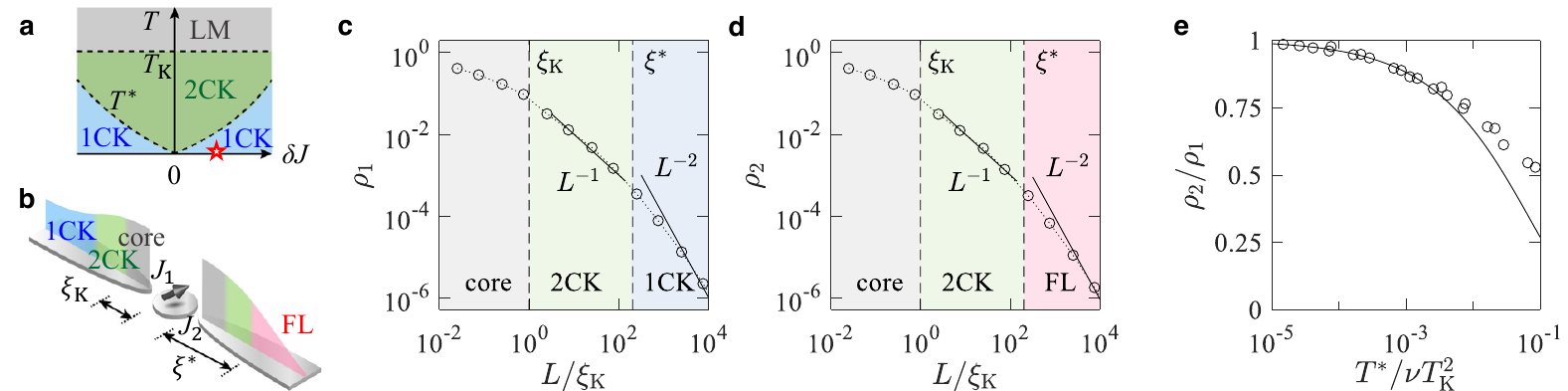}}
\caption{
{\bf Channel-anisotropic two-channel cloud shells.} {\bf a} Two-channel Kondo (2CK) phase diagram.
\DHK{It consists of the local moment (LM), single-channel Kondo (1CK), and two-channel Kondo (2CK) phases. 
$\delta J$ is the channel anisotropy, $T$ is the temperature, $T_\mathrm{K}$ is the Kondo temperature, and $T^*$ is the crossover temperature.}
{\bf b} Cloud distribution at a \DHK{point marked by the red star in the 1CK domain of the phase diagram of {\bf a}. The coupling strengths are $J_{1} = J + \delta J$ and $J_{2} = J - \delta J$. The cloud distribution has the core, 1CK, 2CK and non-Kondo Fermi liquid (FL) shells.} 
$\xi_\mathrm{K}$ is the Kondo length and $\xi^*$ is the crossover length.
{\bf c}, {\bf d} Log-log plots of the distribution $\rho_i (L)$ in channel $i=1,2$ at zero temperature. Cloud shells are identified by their power-law decay.
{\bf e} Ratio $\rho_{2} / \rho_{1}$  at $L \gg \xi^*$.
\DHK{Here $\nu$ is the local density of states.}
The numerical renormalization group (NRG) results (dots) agree with the bosonization prediction (solid curve).
}
\label{MCKC_fig2}
\end{figure*}
\end{center}
\twocolumngrid
 
In Fig.~\ref{MCKC_fig1}, the spatial distribution $\rho_n$ of the entanglement is obtained at zero temperature.
The distribution extends over the whole space, having the core and the tail inside and outside the cloud length $\xi_\textrm{K} = \hbar v / (k_\textrm{B} T_\mathrm{K})$, where $v$ is the Fermi velocity.
$\rho_{n}$ is much larger in the core than in the tail, showing that most electrons forming the cloud lies in the core.
The core does not show any characteristics of the zero-temperature bulk criticality, strongly ``binding'' with the impurity.
The core corresponds to the LM phase~\cite{Mitchell11}.
By contrast, the tail slowly decays, following the universal power law 
\begin{align}
\rho_n (L) \propto \big( \frac{\xi_\mathrm{K}}{L} \big)^{2\Delta} \quad \quad  L \gg \xi_\mathrm{K}. \label{powerlaw}
\end{align} 
We derive Eq.~\eqref{powerlaw} using  the BCFT (Supplementary Note 5), with focusing on the envelope of the Friedel oscillations in the $L$ dependence of $\rho_n$.
The power-law exponent is governed by the scaling dimension $\Delta$ of the BCFT operator describing the impurity spin.
For $k=1$, $\Delta = 1$ which implies the FL of the 1CK.
For $k \ge 2$, $\Delta = 2 / (2+k)$ which signifies the NFL of the $k$CK~\cite{Affleck91b}.
The tail accords with the bulk criticality.
The exponent $\Delta$ at each phase is summarized in Table~\ref{table1}.

The core and tail structure of the entanglement distribution $\rho_n$ is a visualization of the quantum coherent Kondo cloud.
The LSB is useful for the visualization.

\begin{table}[b]
	\begin{center}
		\begin{tabular}{|m{0.8cm} |m{1.0cm} |m{1.0cm} |m{1.0cm} |m{1.6cm}| m{2.1cm}|}
			\hline
		\DHK{shell} & \centering 1CK & \centering 2CK & \centering 3CK & \centering $k(\geq 4)$CK & non-Kondo FL  \\ 
			\hline
			\centering $\Delta$ & \centering $1$ & \centering $1/2$ & \centering $2/5$ & \centering  $2/(k+2)$ &  \quad \quad \quad $1$  \\ \hline
		\end{tabular}
		\caption{{\bf Scaling exponent of cloud shells.}
		Scaling exponent $\Delta$ of the cloud distribution $\rho_n$ of channel $n$ in the single-channel Kondo (1CK), two-channel Kondo (2CK), three-channel Kondo (3CK), $k$-channel Kondo ($k$CK; $k$ is number of channels), and non-Kondo Fermi liquid (FL) shells.}
		\label{table1}
	\end{center}
\end{table}

\begin{center}
\begin{figure*}[t]
\centerline{\includegraphics[width=.9\textwidth]{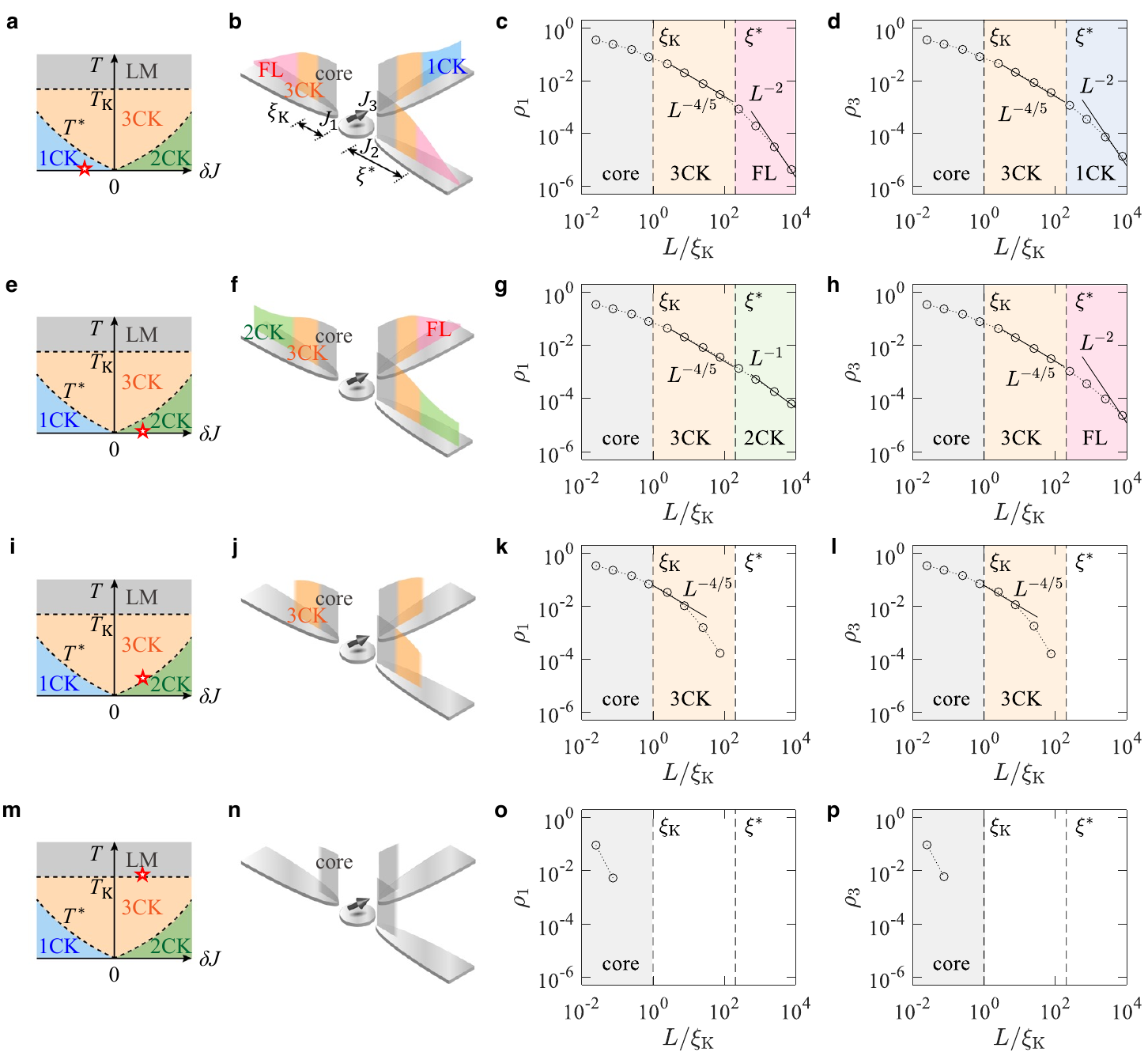}}
\caption{
{\bf Three-channel cloud shells and their thermal evaporation.} The three-channel Kondo (3CK) model of \DHK{couplings} $J_{1,2} = J + (\delta J) / 2$ and $J_{3} = J - \delta J$ is considered.
\DHK{$\delta J$ is the channel anisotropy.}
{\bf a}-{\bf d}
\DHK{The phase diagram of the model, shown in {\bf a}, is composed of the local moment (LM), single-channel Kondo (1CK), two-channel Kondo (2CK), and three-channel Kondo (3CK) phases. At a point of $\delta J < 0$ and zero temperature $T=0$ marked by the red star in the  phase diagram {\bf a}}, the cloud distribution is drawn in {\bf b}, the log-log plot of numerical renormalization group (NRG) results of the distribution $\rho_1 (L)$ is in {\bf c}, and the log-log plot of $\rho_3 (L)$ is in {\bf d}.
$\rho_2$ is identical to $\rho_1$.
\DHK{In {\bf b}, the core, 1CK, 3CK, and non-Kondo Fermi liquid (FL) shells are identified. $T_\mathrm{K}$ is the Kondo temperature, $T^*$ is the crossover temperature, $\xi_\mathrm{K}$ is the Kondo length, and $\xi^*$ is the crossover length.}
{\bf e}-{\bf h} The same plots, but at a point of $\delta J > 0$ and $T=0$.
{\bf i}-{\bf l} The same plots, but at a point of $\delta J > 0$ and $T=T^*$.
{\bf m}-{\bf p} The same plots, but at a point of $\delta J > 0$ and $T=T_\mathrm{K}$. As temperature increases, the outer shells disappear one by one. 
}
\label{MCKC_fig3}
\end{figure*}
\end{center}

 \textbf{Entanglement shells of anisotropic multichannel Kondo clouds ---} 
We next consider channel anisotropic cases of $k$ channels. 
It is known that there are multiple crossover temperatures~\cite{Nozieres80}.
At $T \gtrsim T_\mathrm{K}$, the LM phase happens.
At $T^* \lesssim T \lesssim T_\mathrm{K}$, the Kondo effect by the $k$ channels ($k$CK) occurs, where $T^*$ is a crossover temperature determined by the anisotropy.
Below $T^*$ there can appear $k'$-channel Kondo effects with $k' < k$. The zero temperature phase is a $k''$CK with $k'' \le k'$ where $k''$ is the number of the channels having the largest coupling. 
These are shown in the phase diagrams of Figs.~\ref{MCKC_fig2}a and \ref{MCKC_fig3}a.

We first discuss the Kondo cloud of the anisotrpic $k$CKs at zero temperature.
We find that the spatial distribution $\rho_n$ has the core and the tail of a shell structure [Figs.~\ref{MCKC_fig2} and \ref{MCKC_fig3}a-h].$\rho_n$ is much larger in the core, which appears over $L \lesssim \xi_\textrm{K}$, than in the tail, as in the isotropic case.
The tail has hierarchical multiple shells of distinct entanglement scaling behaviors.
In the innermost shell, all the $k$ channels follow the power law decay of $\rho_n(L) \propto (\xi_\textrm{K} / L)^{2\Delta}$ with $\Delta = 2 / (2+k)$.
This shell corresponds to the NFL of the isotropic $k$CK, as identified by Eq.~\eqref{powerlaw} and shown in Table~\ref{table1},
and appears at $\xi_\textrm{K} \lesssim L \lesssim \xi^*$ with $\xi^* = \hbar v / (k_\textrm{B} T^*)$. The core and the innermost shell are identical between the channels, although the coupling strengths $J_i$ are different.

\onecolumngrid
\twocolumngrid

On the other hand, the other shells are channel dependent.
In the outermost shell, the $k''$ channels having the same coupling strength but larger than the others show different behavior from the others. These largest-coupling channels exhibit the distribution $\rho_n(L)$ of the power law decay with $\Delta = 1$ for $k''=1$ (namely when one channel has stronger coupling than all the others)
and $\Delta = 2 / (2 + k'')$ for $k'' \ge 2$. These channels in the shell exhibit the zero-temperature $k''$CK phase, as implied by Eq.~\eqref{powerlaw}(see also Table~\ref{table1}).
The other $k-k''$ channels of weaker coupling in this shell also have nonzero distribution $\rho_n$, albeit smaller than that of the $k''$ channels.
They follow the power law decay of $\rho_n(L)$ with $\Delta = 1$, showing a non-Kondo FL that does not show the Kondo effect as discussed below.
Hence the outermost shell of the Kondo cloud is composed of the NFL (resp. FL) of the $k''$CK in the $k''$ channels of the strongest coupling for $k'' \ge 2$ (resp. $k''=1$) 
and the non-Kondo FL in the other channels.

We discuss about the non-Kondo FL behavior in the $k-k''$ channels of weaker coupling. The value of $\Delta = 1$ implies that these channels are Fermi liquids.  Although the value is identical to that of the 1CK case (see Table~\ref{table1}), these channels of weaker coupling do not exhibit Kondo behaviors. For example, in an anisotropic 2CK model~\cite{Nozieres78a,Nozieres78b,Carmi12,Mitchell12}, the channel of stronger coupling exhibits the $\pi$ scattering phase shift as in the 1CK case, while the weaker-coupling channel does not. It is interesting that a spin cloud, having an algebraic tail (indicated by the non-vanishing entanglement between the impurity and the channels), is developed in these weaker-coupling channels. A recent work~\cite{Moca21} reported a similar finding that a spin cloud appears in a non-Kondo phase of a superconductor coupled with a magnetic impurity. 

In Figs.~\ref{MCKC_fig2} and \ref{MCKC_fig3}a-h,
these features of the outermost shell are shown for the 2CK of 
$J_{1} = J + \delta J$ and $J_{2} = J - \delta J$, and the 3CK of 
$J_{1,2} = J + (\delta J) / 2$ and $J_{3} = J - \delta J$.
The shell appears at $L \gtrsim \xi^*$,
where 
$\xi^* \propto |\delta J|^{-2} T_\mathrm{K}^{-1}$ for the 2CK and 
$\xi^* \propto |\delta J|^{-5/2} T_\mathrm{K}^{-1}$ for the 3CK~\cite{Iftikhar18}.
At $L \gtrsim \xi^*$ in the 2CK, the channel 1 of stronger coupling has the 1CK FL, while the channel 2 has a non-Kondo FL.
We find, using the bosonization~\cite{Emery92,Zarand00} (Supplementary Note 6), that
the channel 2 shows nonzero distribution $\rho_2$ smaller than the channel 1,
following $\rho_2 /\rho_1 \cong T^*/\nu T_\mathrm{K}^2$ at $L \gg \xi^*$ [Fig.~\ref{MCKC_fig2}e]. $\nu$ is the density of states.
In the 3CK with $\delta J > 0$, the channels 1 and 2 having the largest coupling exhibit the 2CK NFL in the outermost shell, while the channel 3 shows a non-Kondo FL. In the 3CK with $\delta J <0$, the channel 3 of the largest coupling shows the 1CK FL in the outermost shell, while the other channels exhibit a non-Kondo FL.

In general anisotropic $k$CKs, there appear intermediate shells corresponding to a $q_1$CK, a $q_2$CK, $\cdots$ (from outer to inner) between the innermost and outermost shells, 
with the hierarchy $k'' < q_1 < q_2 < \cdots < k$ determined by the coupling strengths $J_{n=1,2,\cdots,k}$. In the shell of the $q_i$CK, the $q_i$ channels having larger coupling than the others exhibit the $q_i$CK NFL, while the other $k-q_i$ channels show a non-Kondo FL. For example, we find that in the most general case of the 3CK with $J_1 > J_2 > J_3$, the Kondo cloud is composed of the core, the innermost 3CK shell, the intermediate 2CK shell (having the 2CK NFL in two channels of larger coupling and a non-Kondo FL in the other), and the outermost 1CK shell (having the 1CK FL in the channel of the largest coupling and a non-Kondo FL in the others) at zero temperature (Supplementary Note 4).


 \textbf{Thermal evaporation of entanglement shells ---} 
To examine the thermal decoherence of the entanglement shells and hence the Kondo cloud, we compute 
$\rho_n (L,T)$ in Eq.~\eqref{eq:profile} at finite temperatures, using the NRG.
$\rho_n (L,T) = \mathcal{N}_0 (T) - \mathcal{N} (L,T; n)$ quantifies the difference of the entanglement between the absence and presence of the LSB at temperature $T$; $\mathcal{N}_0 (T)$ measures the entanglement that survives against thermal fluctuations at $T$, while $\mathcal{N} (L,T; n)$ measures the entanglement at $T$ further reduced by the LSB at distance $L$ in channel $n$. More reduction occurs as the impurity spin is more entangled with (i.e., more screened by) electrons at $L$. Hence, $\rho_n (L,T)$ quantifies the entanglement distribution at $T$ with varying $L$.
Note that in the absence of the LSB, the entanglement algebraically decays thermally~\cite{Kim21}, $\mathcal{N}_0 (T) = 1 - a_k (T/T_\textrm{K})^{2 \Delta}$ at $T \ll T_\textrm{K}$, where $a_k > 0$ is a constant. 
 
For the 3CK with $\delta J >0$, Figs.~\ref{MCKC_fig3}e-p show the temperature dependence of the entanglement shells. 
Thermal fluctuations suppress shells outside the thermal length $\hbar v / (k_\textrm{B} T)$, while it does almost not affect shells inside. So the outer shells are thermally ``evaporated'' one by one. 
At $T \ll T^*$, the outermost shell, located at $L > \xi^*$, shows the 2CK NFL in the channels 1 and 2, as discussed above.
At $T^* \lesssim T \lesssim T_\textrm{K}$, the outermost shell is almost suppressed. Then the remaining inner shell at $\xi_\textrm{K} \lesssim L \lesssim \xi^*$, whose character is the 3CK NFL, determines the thermal phase. When the temperature further increases to $T \gtrsim T_\textrm{K}$, only the core at $L \lesssim \xi_\textrm{K}$ survives and represents the LM thermal phase.

This clearly shows that the hierarchical shells of the boundary entanglement at zero temperature is the manifestation of the renormalization group flow in the development of the Kondo effects.
Inner shells are ``bound'' more strongly with, namely more entangled with, the impurity, being more robust against thermal fluctuations. 
Namely, inner shells cause the boundary condition of the bulk conduction electrons of higher energies, hence, determining phases at higher temperature.
Note that a related temperature dependence of a single-channel Anderson impurity model was discussed in Ref.~\cite{Mitchell11}.


 \textbf{How to detect boundary entanglement shells ---} 
Equation~\eqref{magnetization_zero} implies that the entanglement distribution $\rho_n(L)$, hence, the Kondo cloud can be experimentally detected by  monitoring the change of the impurity magnetization with varying the position $L$ of an LSB in a channel $n$.
The relation is exact at zero temperature and a very good approximation at $T \ll T_\textrm{K}$ and $L \lesssim \hbar v / (k_\textrm{B} T)$ where thermal fluctuations negligibly affect $\rho_n(L)$ as demonstrated in Fig.~\ref{MCKC_fig3}.

\begin{figure}[t]
	\centerline{\includegraphics[width=.9\columnwidth]{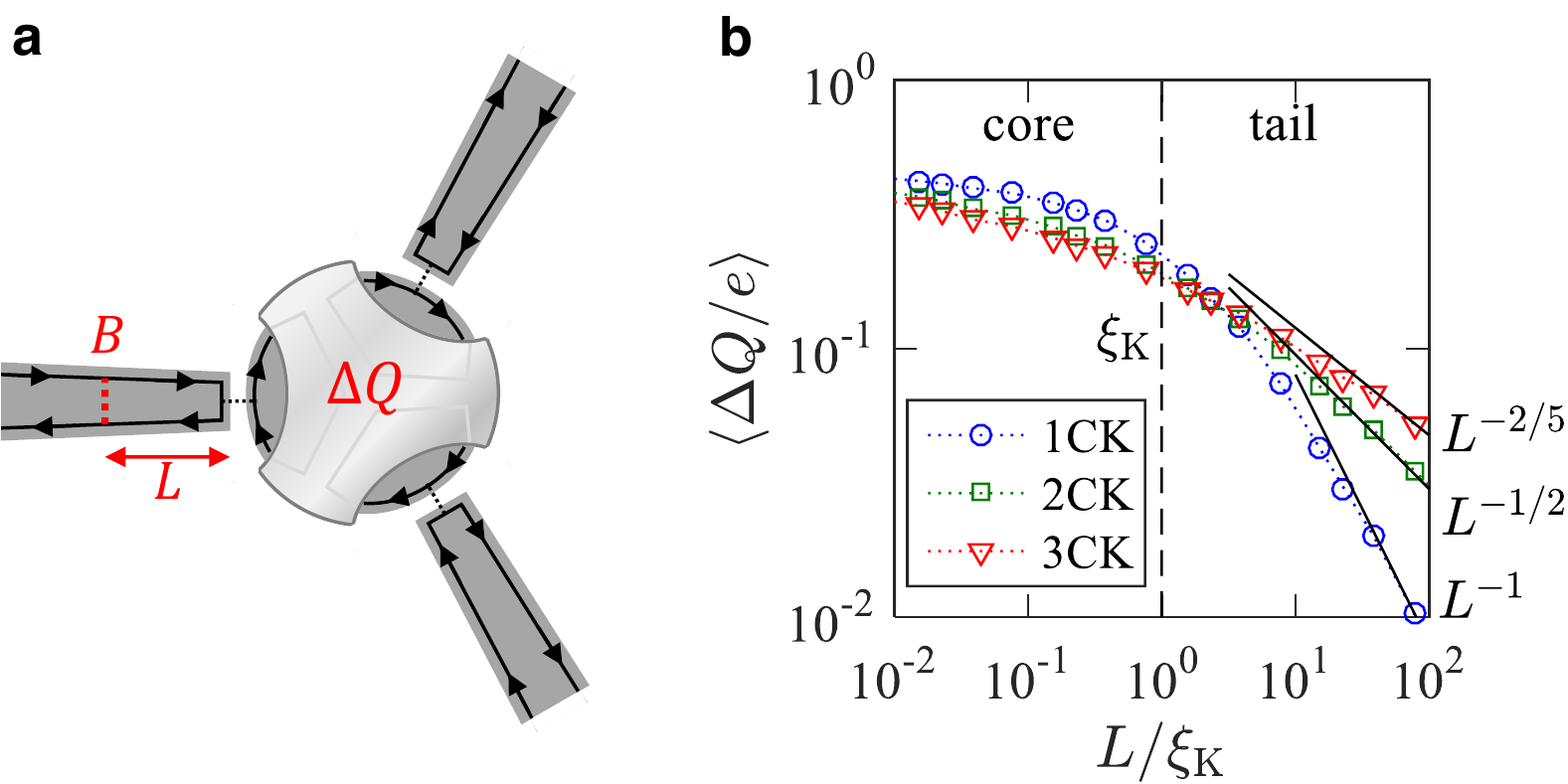}}
	\caption{{\bf How to detect Kondo clouds.} {\bf a} A metallic dot couples with quantum Hall edge channels.
		Its excess charge $\Delta Q$ supports a Kondo impurity pseudospin. 
		The cloud is formed in the channels. 
		An local symmetry breaking (LSB) \DHK{of strength $B$} is applied by placing a quantum point contact in a channel $n$ at distance $L$ from the dot.
		{\bf b} Numerical renormalization group (NRG) results of $\langle \Delta Q \rangle$ as a function of $L$ for \DHK{the isotropic single-channel Kondo (1CK), two-channel Kondo (2CK), and three-channel Kondo (3CK)} effects.
		\DHK{$e$ is the electron charge and  $\xi_\mathrm{K}$ is the Kondo length.}
	}
	\label{MCKC_fig4}
\end{figure}

We propose an experiment based on a charge-Kondo circuit~\cite{Iftikhar15,Iftikhar18} with which multichannel Kondo effects can be manipulated.  It has a metallic dot coupled to $k$ quantum Hall edge channels (Fig.~\ref{MCKC_fig4}). 
Energy-degenerate charge states $|N \rangle$ and $|N+1 \rangle$ of the dot form the pseudospin 1/2,
and the excess charge $\Delta Q \equiv Q - (N + 1/2) e$ of the dot plays the role of the magnetization $M/\hbar$ of the pseudospin. 
Here $N$ and $Q$ denote the number of electrons and the charge operator for the dot, respectively,
and $e$ is the electron charge.

We show that a quantum point contact placed on a channel $n$ at distance $L$ from the dot results in an LSB breaking the SU(2) pseudospin symmetry (Fig.~\ref{MCKC_fig4}, Supplementary Note 7).
At $T=0$, 
the negativity in the absence of the LSB is $\mathcal{N}_{0}(T = 0)= 1$~\cite{Kim21}, while the negativity  in the presence of the LSB is $\mathcal{N}(L,T = 0;n) = \sqrt{1-4\langle\Delta Q/e\rangle^2}$ [see Eq.~\eqref{magnetization_zero}]. 
These give $\rho_n(L,T = 0) = \mathcal{N}_{0}(T = 0) - \mathcal{N}(L,T = 0;n) = 1 -\sqrt{1-4\langle\Delta Q/e\rangle^2} \simeq 2\langle\Delta Q / e \rangle^2$ for small $\langle\Delta Q / e \rangle \ll 1$.
At low temperature 
$T \ll T_{\mathrm{K}}$ where thermal fluctuation on $\langle\Delta Q / e \rangle^{2}$ is negligible,
$\rho_n(L,T)$ can be approximated as the zero-temperature value of $\rho_{n}(L,T = 0) \simeq 2\langle\Delta Q / e \rangle^{2}$.
It is possible to measure $\Delta Q (L)$, hence $\rho_n(L)$, by monitoring electric current through another quantum point contact~\cite{Field93} nearby the dot.
The entanglement shells in isotropic and anisotropic $k$CKs can be experimentally identified with realistic parameters (Supplementary Note 7).

\begin{center}
\textbf{Discussion}
\end{center}

Our work demonstrates how a spin cloud screening a local magnetic impurity in a metal differs at the fundamental level from a charge cloud screening an excess charge. 
For the demonstration, we developed a theory of the boundary-bulk entanglement in multichannel Kondo effects. Utilizing an LSB, the spatial distribution and thermal suppression of the entanglement can be computed and experimentally detected. The distribution is a visualization of the spatial and energy structure of the quantum coherent Kondo spin screening cloud. 

The boundary-bulk entanglement is applicable to general boundary quantum critical phenomena as below.
The entanglement quantifies the quantum coherent coupling between the boundary and the bulk in boundary criticalities.
Its spatial structure will have information of competing phases or boundary conditions, as suggested by the hierarchical shells of Kondo clouds.
In spin-1/2 boundary criticalities, it is obtained, using the boundary magnetization and Eq.~\eqref{magnetization_zero}. 
In more general cases, it may be calculated with BCFT boundary operators~\cite{Kim21}.

An LSB that breaks the boundary-bulk coupling symmetry will be useful for identifying the boundary structure of boundary criticalities. 
The spatial structure is estimated by the change of the entanglement as a function of the location of the LSB, while the partition for the entanglement is placed at the boundary.
This differs from the usual way~\cite{Affleck09a}
where an entanglement is studied with placing the entanglement partition in the bulk.
 
The boundary-bulk entanglement will be experimentally accessible.
As in Eq.~\eqref{magnetization_zero}, it may have a simple relation with a boundary observable, when the entanglement has a simple form like Kondo singlets near a fixed point of boundary criticalities.
Such a simple relation between an entanglement and an observable is rare.
It is another usefulness of the boundary-bulk entanglement.

We anticipate that the boundary-bulk entanglement is an essential aspect of 
boundary criticalities and related effects such as Kondo lattices and heavy fermions~\cite{Si01,Coleman15,Georges96}.
 



\begin{center}
\textbf{Data availability}
\end{center}
All the calculation details are provided in Supplementary Information.


\subsection{Acknowledgements}
We thank Frederic Pierre and Jan von Delft for useful discussions. This work is supported by Korea NRF via the SRC Center for Quantum Coherence in Condensed Matter (Grant No. 2016R1A5A1008184 and RS-2023-00207732) and Grant No. 2023R1A2C2003430. D.K. acknowledges support by Korea NRF via Basic Science Research Program for Ph.D. students (2022R1A6A3A13062095).

\subsection{Author Contributions}
J.S. performed the NRG. D.K. performed the BCFT and bosonization calculation. H.-S.S. supervised the project. All the authors were involved in developing the theoretical approach and preparing the manuscript.

\subsection{Competing interests}
The authors declare no competing interests.



\clearpage
\newpage
\setcounter{page}{1}

\onecolumngrid

\begin{center}
\textbf{\large Supplementary Information for \textquotedblleft Hierarchical entanglement shells of multichannel Kondo clouds \textquotedblright}
\end{center}

\begin{center}
Jeongmin Shim$^{1,2,3}$, Donghoon Kim$^{1,3}$, and H.-S. Sim$^{1,*}$

\textit{\small $^{1}$Department of Physics, Korea Advanced Institute of Science and Technology, Daejeon 34141, Korea} \\
\textit{\small $^{2}$Present address: Arnold Sommerfeld Center for Theoretical Physics, Center for NanoScience, and Munich Center for Quantum Science and Technology, Ludwig-Maximilians-Universit\"at M\"unchen, 80333 Munich, Germany}\\
\textit{\small $^{3}$These authors contributed equally: Jeongmin Shim, Donghoon Kim}
\end{center}

\newcommand{\beginsup}{
        \setcounter{equation}{0}
        \renewcommand{\theequation}{S\arabic{equation}}
        \setcounter{table}{0}
        \renewcommand{\thetable}{\arabic{table}}
        \setcounter{figure}{0}
        \renewcommand{\figurename}{Supplementary Fig.}
     }
\beginsup

This material contains details in the NRG, BCFT, and bosonization calculations and estimate of experimental parameters. Below we set $\hbar \equiv 1$, $v_\textrm{F} = 1$, and $k_\textrm{B} \equiv 1$ somewhere for simplicity.

\begin{center}
\textbf{Supplementary Note 1. DERIVATION OF EQ.~(3)}
\end{center}
We derive Eq.~(3) of the main text. An equivalent proof is found in Supplementary Materials (see Sec.~III) of Ref.~\cite{Kim21_supp}. We consider the $k$CK model affected by the LSB, i.e., $H = H_{k\mathrm{CK}} + H_{\mathrm{LSB}}$ [See Eqs.~(1) and (4) of the main text]. In this case, the $\mathrm{SU}(2)$ symmetry is broken and the ground state becomes nondegenerate. 
The nondegenerate ground state $|E \rangle$ is written in a Schmidt decomposed form,
\begin{align}\label{Schmidt1}
|E \rangle = \sum_{i=1,2} \sqrt{p_{i}} |\phi_{i} \rangle_{A} \otimes |\psi_{i} \rangle_{B},
\end{align}
where $A$ and $B$ denote the impurity and its environment, respectively, and $p_{1} + p_{2} = 1$. Since the Hilbert-space dimension of the impurity is 2, the state is expressed by two orthonormal bases $\{|\phi_{i} \rangle_{A}\}_{i=1,2}$ and $\{|\psi_{i} \rangle_{B}\}_{i=1,2}$. 
In the basis $\{|\phi_{1} \rangle_{A} |\psi_{1} \rangle_{B}, |\phi_{1} \rangle_{A} |\psi_{2} \rangle_{B}, |\phi_{2} \rangle_{A} |\psi_{1} \rangle_{B}, |\phi_{2} \rangle_{A} |\psi_{2} \rangle_{B}\}$, the density matrix and its partial transpose are written as  
\begin{align}
\rho = |E \rangle \langle E| \doteq \begin{pmatrix} p_{1} & 0 & 0 & \sqrt{p_{1} p_{2}} \\ 0 & 0 & 0 & 0 \\ 0 & 0 & 0 & 0 \\ \sqrt{p_{1} p_{2}} & 0 & 0 & p_{2} \end{pmatrix}, \qquad \rho^{\mathrm{T}_{A}} \doteq \begin{pmatrix} p_{1} & 0 & 0 & 0 \\ 0 & 0 & \sqrt{p_{1} p_{2}} & 0 \\ 0 & \sqrt{p_{1} p_{2}} & 0 & 0 \\ 0 & 0 & 0 & p_{2} \end{pmatrix}.
\end{align}
Here, $\mathrm{T}_{A}$ is the partial transpose on the impurity. Then the singular values of $\rho^{\mathrm{T}_{A}}$ are $p_{1}$, $p_{2}$, and two $\sqrt{p_{1}p_{2}}$, so the trace norm $\Vert \rho^{\mathrm{T}_{A}} \Vert_{1}$, the sum of these singular values, is $p_{1} + p_{2} + 2 \sqrt{p_{1} p_{2}} = 1 + 2 \sqrt{p_{1} p_{2}}$. Therefore, it leads to the entanglement negativity between the impurity and its enviroment
\begin{align}\label{Negativity1}
\mathcal{N} = \Vert \rho^{\mathrm{T}_{A}} \Vert_{1} - 1 = 2 \sqrt{p_{1} p_{2}},
\end{align}
which is solely determined by $p_{1}$ and $p_{2}$.

We now calculate $p_{1}$ and $p_{2}$. From Eq.~\eqref{Schmidt1}, the reduced density matrix of the impurity system is derived as
\begin{align}\label{RDM}
\rho_{A} = \mathrm{Tr}_{B}[|E \rangle \langle E|] = \sum_{i=1,2} p_{i} |\phi_{i} \rangle_{A} \langle \phi_{i}|.
\end{align}
Namely, $\{|\phi_{i} \rangle_{A}\}_{i=1,2}$ and $\{p_{i}\}_{i=1,2}$ are the eigenvectors and eigenvalues of $\rho_{A}$. Since $\rho_{A}$ is an operator on the impurity that has two levels, it is written as a linear combination of the identity operator $\mathds{I}$ and Pauli matrices $\sigma_{x,y,z}$,
\begin{align}\label{RDM2}
\rho_{A} = a_{0} \mathds{I} + a_{1} \sigma_{x} + a_{2} \sigma_{y} + a_{3} \sigma_{z}. 
\end{align}
Using $\mathrm{Tr}[\rho_{A}] = 1$ and $\mathrm{Tr}[\rho_{A} S_{\mathrm{imp}}^{x,y,z}] = \langle E| S_{\mathrm{imp}}^{x,y,z} |E \rangle \equiv \langle S_{\mathrm{imp}}^{x,y,z} \rangle$ ($S_{\mathrm{imp}}^{\alpha} = \hslash \sigma_{\alpha} / 2$ is the impurity spin for $\alpha = x,y,z$), Eq.~\eqref{RDM2} becomes
\begin{align}\label{RDM3}
\rho_{A} = \frac{1}{2} \mathds{I} + \frac{\langle S_{\mathrm{imp}}^{x} \rangle}{\hslash} \sigma_{x} + \frac{\langle S_{\mathrm{imp}}^{y} \rangle}{\hslash} \sigma_{y} + \frac{\langle S_{\mathrm{imp}}^{z} \rangle}{\hslash} \sigma_{z}. 
\end{align}
The eigenvalues of $\rho_{A}$ in Eq.~\eqref{RDM3} are
\begin{align}\label{RDMEig}
p_{1} = \frac{1}{2} \Bigg(1 + \frac{2}{\hslash} \sqrt{\langle S_{\mathrm{imp}}^{x} \rangle^{2} + \langle S_{\mathrm{imp}}^{y} \rangle^{2} + \langle S_{\mathrm{imp}}^{z} \rangle^{2}}\Bigg), \qquad p_{2} = \frac{1}{2} \Bigg(1 - \frac{2}{\hslash} \sqrt{\langle S_{\mathrm{imp}}^{x} \rangle^{2} + \langle S_{\mathrm{imp}}^{y} \rangle^{2} + \langle S_{\mathrm{imp}}^{z} \rangle^{2}}\Bigg).
\end{align}
Combining Eqs.~\eqref{Negativity1} and \eqref{RDMEig}, we  obtain
\begin{align}\label{Negativity2}
\mathcal{N} = \sqrt{1 - \frac{4}{\hslash^{2}} (\langle S_{\mathrm{imp}}^{x} \rangle^{2} + \langle S_{\mathrm{imp}}^{y} \rangle^{2} + \langle S_{\mathrm{imp}}^{z} \rangle^{2})}.
\end{align}
Since $\mathbf{M} = \langle E | \mathbf{S}_{\mathrm{imp}} |E \rangle = \langle \mathbf{S}_{\mathrm{imp}} \rangle = \langle S_{\mathrm{imp}}^{x} \rangle \hat{\mathbf{e}}_{x} + \langle S_{\mathrm{imp}}^{y} \rangle \hat{\mathbf{e}}_{y} + \langle S_{\mathrm{imp}}^{z} \rangle \hat{\mathbf{e}}_{z}$, 
Eq.~\eqref{Negativity2} is equal to Eq.~(3) of the main text.

\begin{center}
\textbf{Supplementary Note 2. NRG CALCULATION}
\end{center}

Our NRG calculation of the entanglement negativity $\mathcal{N}$ is done, using the method developed in Ref.~\cite{Shim18_supp}. Below we describe the Hamiltonian and parameters used in the calculation.

In the total Hamiltonian $H_{k\mathrm{CK}} + H_\mathrm{LSB}$, $H_{k\mathrm{CK}}$ describes the $k$CK model,
\begin{align}
H_{k\mathrm{CK}} &= \sum_{j=1}^k H_j + \sum_{j=1}^k  J_j \mathbf{S}_\mathrm{imp} \cdot \mathbf{S}_j .
\end{align}
$H_j = \frac{D}{2}\sum_{\alpha=\pm} \sum_{\ell=0} \big[\psi_{\alpha j}^{\dagger}(\ell) \psi_{\alpha j}(\ell + 1) + \psi_{\alpha j}^{\dagger}(\ell + 1) \psi_{\alpha j}(\ell)\big]$ is a semi-infinite one-dimensional tight-binding chain Hamiltonian for the $j$th conduction channel. $D$ is  a half band width  of the chain. $\psi_{\alpha j}(\ell)$ is an annihilation operator of an electron having spin $\alpha$ at the $\ell$-th site of the $j$th chain. $\alpha = \pm$ represents the eigenspin states of the spin operator  $S^x$  in the $x$ direction. The impurity spin $\mathbf{S}_\mathrm{imp}$ is coupled to the spin $\mathbf{S}_j$ at the $0$-th site of the $j$th chain with the coupling strength $J_j$. 
The local spin symmetry breaking perturbation $H_\mathrm{LSB}$ at $n$th conduction channel is described by
\begin{align}
H_\mathrm{LSB} &= \frac{B}{2} \big[\psi_{+ n}^{\dagger}(L) \psi_{+ n}(L) - \psi_{- n}^{\dagger}(L) \psi_{- n}(L)\big] . \label{LSBxdirection}
\end{align}
This Hamiltonian breaks the SU(2) spin symmetry such that an electron has a different energy depending on the direction of its spin  $S^x$  at the $L$-th site of the $n$th channel. $B$ is the strength of the symmetry breaking.

To solve the total Hamiltonian $H_{k\mathrm{CK}} + H_\mathrm{LSB}$  in the NRG approach,  
we obtain the local densities of states (LDOSs) of conduction electrons at the $0$-th site of the channels,
based on the Hamiltonian $\sum_{j=1}^k H_j  + H_\mathrm{LSB}$  of conduction electrons
and using the equations of motion for the Green function $G(\epsilon)$. 
In the  $j$($\neq n$)-th channel where the local symmetry breaking perturbation is not applied,
the Green function of an electron of spin $\alpha$ and energy $\epsilon$ at the $0$-th site is $G_{\alpha j}= G_0 =2(\epsilon-i\sqrt{D^2-\epsilon^2})/D^2$ and the LDOS is found as $\nu_{\alpha j} = -\frac{1}{\pi}\mathrm{Im}[G_{\alpha j}(\epsilon)]$.
In the $n$th channel where the local symmetry breaking perturbation is applied, we solve the equations of motion and find that 
the Green function of an electron of spin $\alpha = \pm$ and energy $\epsilon$ at the $0$-th site is
\begin{align}
G_{\alpha n} (\epsilon) = 
\frac{\sqrt{D^2-\epsilon^2}+(2/\tilde{G}-\epsilon)\mathrm{tan}(L\varphi)}
{\sqrt{D^2-\epsilon^2}+(\epsilon-\tilde{G}D^2/2)\mathrm{tan}(L\varphi)}\tilde{G} ,
\end{align}
where $\tilde{G} (\epsilon)= [(\epsilon + \alpha B) - (D/2)^2 G_0 (\epsilon) ]^{-1}$ and $\varphi=\mathrm{atan}(\sqrt{(D/\epsilon)^2-1})$. Then the LDOS is obtained as $\nu_{\alpha n} = -\frac{1}{\pi}\mathrm{Im}[G_{\alpha n}(\epsilon)]$.
When the half band width $D$ is much larger than temperature $T$ and the  strength $B$ of the spin symmetry breaking, 
the LDOSs approximately follow $\nu_{\alpha j} \sim 1/2D$ for $j \neq n$ and
\begin{align}\label{LDOS}
\nu_{\alpha n}(\epsilon) \sim \frac{1}{2D} \bigg[1 - \alpha (-1)^L \frac{2B}{D} \mathrm{sin}\Big(\frac{2\epsilon L}{D}\Big) \bigg] .
\end{align}
We use these approximated LDOSs in the NRG calculation.
 
We discuss the details~\cite{Wilson75_supp,Bulla08_supp} and the parameters of the NRG calculation.
We employ the full density matrix NRG~\cite{Weichselbaum07_supp,Weichselbaum12_supp} and the interleaved NRG~\cite{Mitchell14_supp,Stadler16_supp} for spin and channel indices.
We set the half band width $D=1$, the discretization parameter $\Lambda=10$, and the length of the Wilson chain by 28.
We choose the number of kept states by 300 for the 1CK model, 3,000 for 2CK, and 10,000 for 3CK.
The result in Fig.~1 of the main text is obtained with the Kondo coupling $J=0.3D$, the perturbation strength $B=0.1D$, and the $z$-averaging of $z=0$, $1/4$, $1/2$ and $3/4$. 
The result in Figs.~2 and 3 is obtained with $J=0.3D$, $B=0.1D$, and the $z$-averaging of $z=0$ and $1/2$.
The result in Fig.~4 is obtained with $J=0.4D$, $B=0.4D$, and the $z$-averaging of of $z=0$ and $1/2$.
We choose the Kondo temperature $T_\mathrm{K}=\mathrm{exp}(-1/\nu J)$ and the Kondo length $\xi_\mathrm{K}=\hbar v_\mathrm{F}/k_\mathrm{B} T_\mathrm{K}$ where $\nu=1/(2D)$.
For the Kondo coupling $J=0.3D$, $T_\mathrm{K}\sim 1.273 \times 10^{-3}D$ and $\xi_\mathrm{K} \sim 4.937 \times 10^{3}/D$. 
We set the Planck constant $h=1$, the Boltzmann constant $k_\mathrm{B}=1$, and the Fermi velocity $v_\mathrm{F}=1$ in the NRG calculation.

We note that the results of $\rho_n < \mathcal{O}(10^{-5})$ and $\mathcal{O}(10^{-3})$ are not shown in Fig.~\ref{MCKC_fig3} of the main text, respectively, as the NRG is less accurate at higher $T$~\cite{Wilson75_supp,Bulla08_supp}.

\newpage
\begin{center}
\textbf{Supplementary Note 3. UNIVERSAL SCALING OF KONDO CLOUD}
\end{center}
In Supplementary Figs.~\ref{MCKC_supp_123}-\ref{MCKC_supp_3to2}, we present the universal scaling of the spatial distribution of Kondo clouds with respect to the Kondo length $\xi_\mathrm{K}$ and the crossover length $\xi^*$.

\begin{figure*}[h]
\centerline{\includegraphics[width=.92\textwidth]{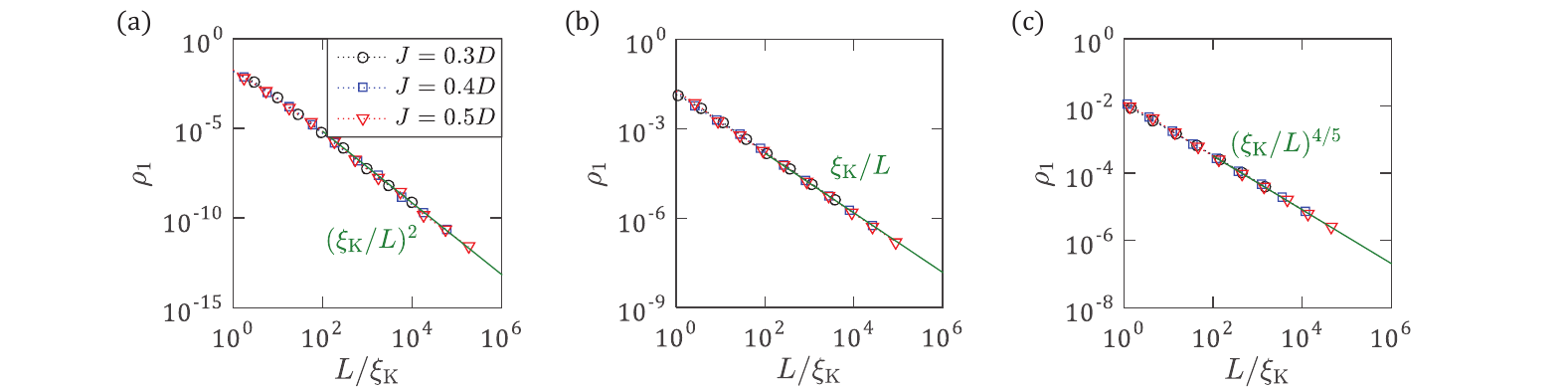}}
\caption{
Universal scaling of the spatial distribution of Kondo clouds.
Distribution $\rho_1$ is drawn as a function of $L/\xi_\mathrm{K}$ for various Kondo coupling $J$ in (a) the single-channel Kondo (1CK), (b) the isotropic two-channel Kondo (2CK), and (c) the isotropic three-channel Kondo (3CK) effects.
Data points from numerical renormalization (NRG) calculations with different values of $J$ lie on a single curve well fitted by  the boundary conformal field theory (BCFT) prediction of $\rho_1 \propto (\xi_\textrm{K} / L)^{2\Delta}$, showing  the universal scaling of the cloud tail in the isotropic Kondo effects.
The distribution on different channels is identical to $\rho_1$ in the isotropic Kondo effects.
In the NRG calculation, we choose $J=0.3D$, $B=0.1D$, and the $z$-averaging of $z=0$ and $1/2$.
}
\label{MCKC_supp_123}
\end{figure*}

\begin{figure*}[h]
\centerline{\includegraphics[width=.92\textwidth]{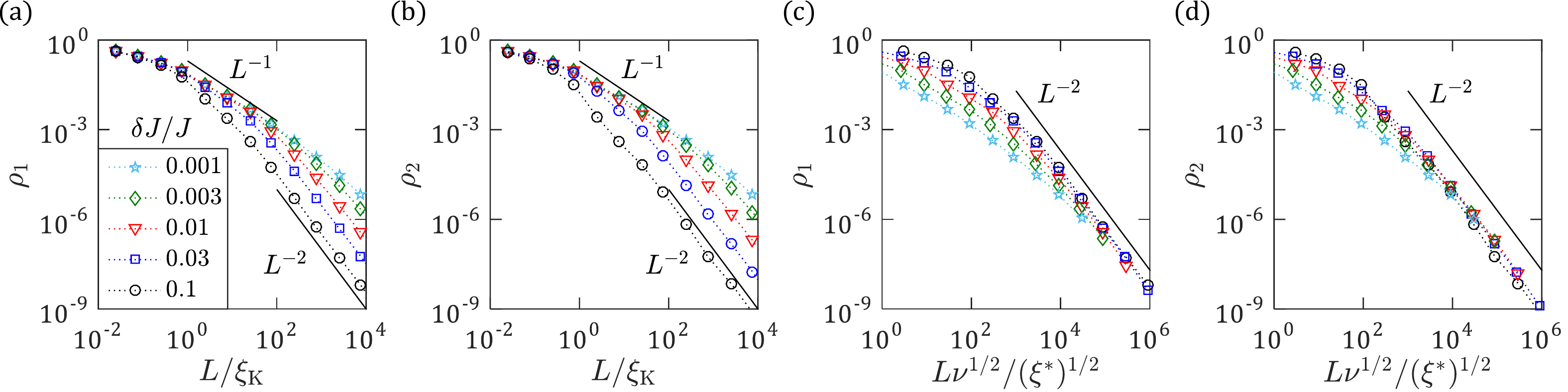}}
\caption{
Universal scaling of the cloud distribution in the anisotropic two-channel Kondo (2CK) effect with $J_1>J_2$.
(a,b) Distribution $\rho_1$ and $\rho_2$ as a function of $L/\xi_\mathrm{K}$ for various channel anisotropy $\delta J$.
Crossover from the 2CK region of $\rho_{1,2} \propto L^{-1}$ to the single-channel Kondo (1CK) region of $\rho_{1} \propto L^{-2}$ and  the non-Kondo Fermi liquid of $\rho_{2} \propto L^{-2}$ is shown. 
(c,d) The results are redrawn as a function of $L \nu^{1/2} /(\xi^*)^{1/2}$.
The numerical renormalization group (NRG) results with various $\delta J$ lie on a single curve at $L > \xi^*$ in agreement with the bosonization prediction of $\rho_{1,2} \propto L^{-2}$.
We choose $J_1=J+\delta J$, $J_2=J-\delta J$, $J=0.4D$, $B=0.1D$, and the $z$-averaging of $z=0$ and $1/2$.
}
\label{MCKC_supp_2to1}
\end{figure*}

\newpage
\begin{figure*}[h]
\centerline{\includegraphics[width=.92\textwidth]{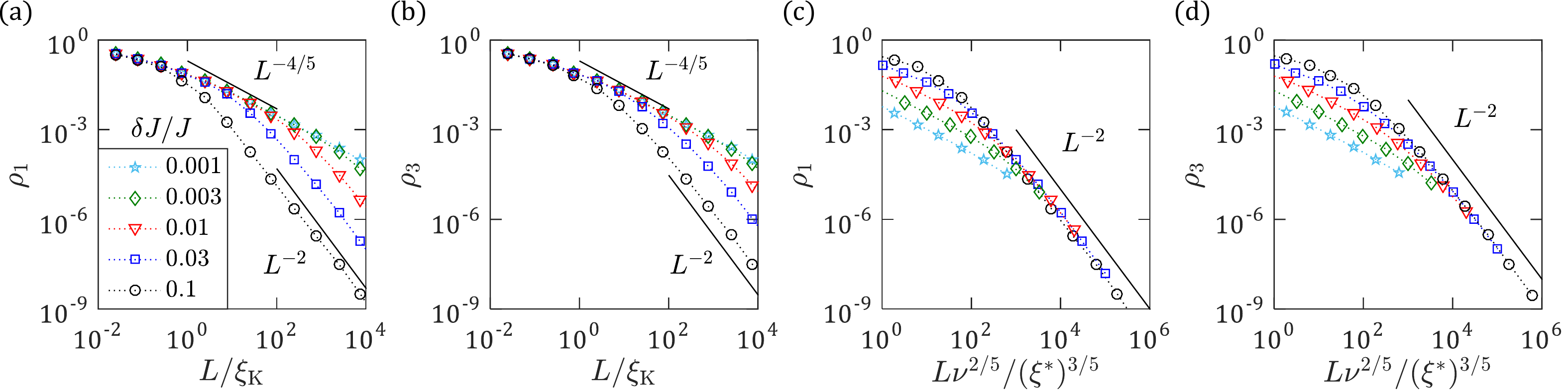}}
\caption{
Universal scaling of the cloud distribution in the anisotropic three-channel Kondo (3CK) effect with  $J_1=J_2<J_3$.
(a,b) Distribution $\rho_1$ and $\rho_3$ as a function of $L/\xi_\mathrm{K}$ for various $\delta J$. Crossover from the 3CK region of $\rho_{1,2,3} \propto L^{-4/5}$ to the single-channel Kondo (1CK) region of $\rho_{3} \propto L^{-2}$ and  the non-Kondo Fermi liquid of $\rho_{1,2} \propto L^{-2}$ is shown. 
(c,d) The results, redrawn as a function of $L \nu^{2/5} /(\xi^*)^{3/5}$, 
lie on a single curve $\propto L^{-2}$. We choose  $J_{1,2}=J-\delta J/2$, $J_3=J+\delta J$, $J=0.4D$, $B=0.1D$, and the $z$-averaging of $z=0$ and $1/2$.
}
\label{MCKC_supp_3to1}
\end{figure*}

\begin{figure*}[h]
\centerline{\includegraphics[width=.92\textwidth]{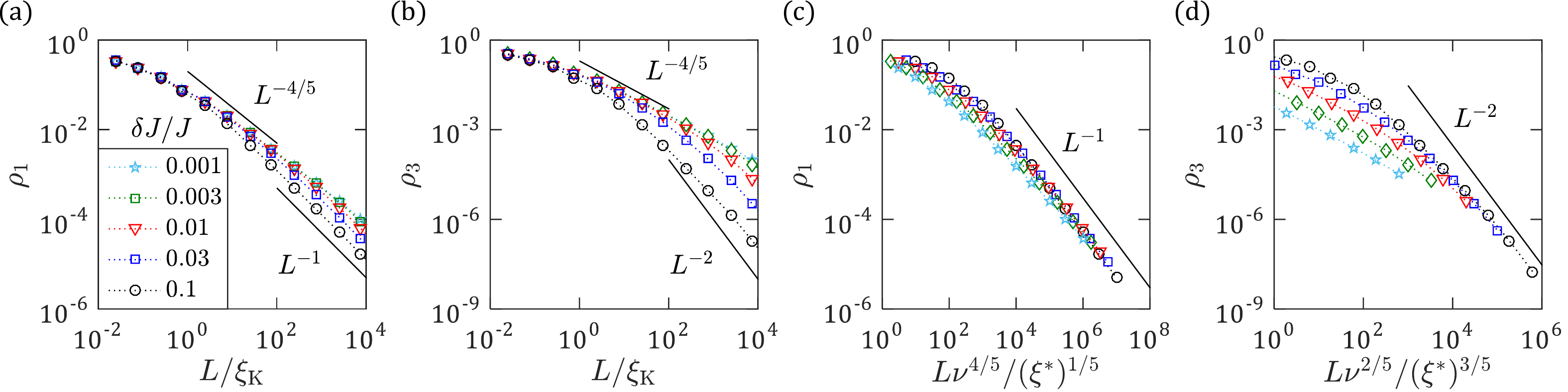}}
\caption{
Universal scaling of the cloud distribution  in the anisotropic three-channel Kondo (3CK) effect with $J_1=J_2>J_3$.
(a,b) Distribution $\rho_1$ and $\rho_3$ as a function of $L/\xi_\mathrm{K}$ for various $\delta J$. 
The distribution $\rho_2$ is equal to $\rho_1$.
Crossover from the 3CK region of $\rho_{1,2,3} \propto L^{-4/5}$ to the two-channel Kondo (2CK) region of $\rho_{1,2} \propto L^{-1}$ and the non-Kondo Fermi liquid of $\rho_{3} \propto L^{-2}$ is shown.
(c,d) The results are redrawn as a function of $L \nu^{4/5} /(\xi^*)^{1/5}$ for $\rho_1$ and $L \nu^{2/5} /(\xi^*)^{3/5}$ for $\rho_3$. 
They lie on a single curve $\propto L^{-1}$ for $\rho_1$ and $\propto L^{-2}$ for $\rho_3$.
We choose $J_{1,2}=J+\delta J/2$, $J_3=J-\delta J$, $J=0.4D$, $B=0.1D$, and the $z$-averaging of $z=0$ and $1/2$.
}
\label{MCKC_supp_3to2}
\end{figure*}

\newpage
\begin{center}
\textbf{Supplementary Note 4. ENTANGLEMENT SHELLS OF KONDO CLOUD WITH GENERAL CHANNEL ANISOTROPY}
\end{center}

We present the spatial distribution of Kondo clouds in the 3CK effects where the coupling strengths all are different,  $J_1 > J_2 > J_3$.
The shell structure of the distribution is found in Supplementary Fig.~\ref{MCKC_supp_123_T}. 
The result shows that the shell structure reflects crossover between different Kondo fixed points.
From the innermost region, there occur four different shells. All the channels have the core region in common, which corresponds to the local moment phase. The core region is followed by the 3CK shell in all the channels.
The next outer shell depends on the relative coupling strengths. When $J_1 > J_2 > J_3$ or $J_1 = J_2 > J_3$, the channels 1 and 2 have the 2CK region in the shell, while the channel 3 has the non-Kondo Fermi liquid region.  When $J_1 > J_2 = J_3$, the channel 1 has the 1CK region in the shell, while the channels 2 and 3 have the non-Kondo Fermi liquid region. This shell is the outermost shell when $J_1 = J_2 > J_3$ or $J_1 > J_2 = J_3$.
When $J_1 > J_2 > J_3$, there occurs another shell in the outmost region, which shows the 1CK region in the channel 1 and the non-Kondo Fermi liquid region in the channels 2 and 3.

\begin{figure*}[h]
\centerline{\includegraphics[width=.98\textwidth]{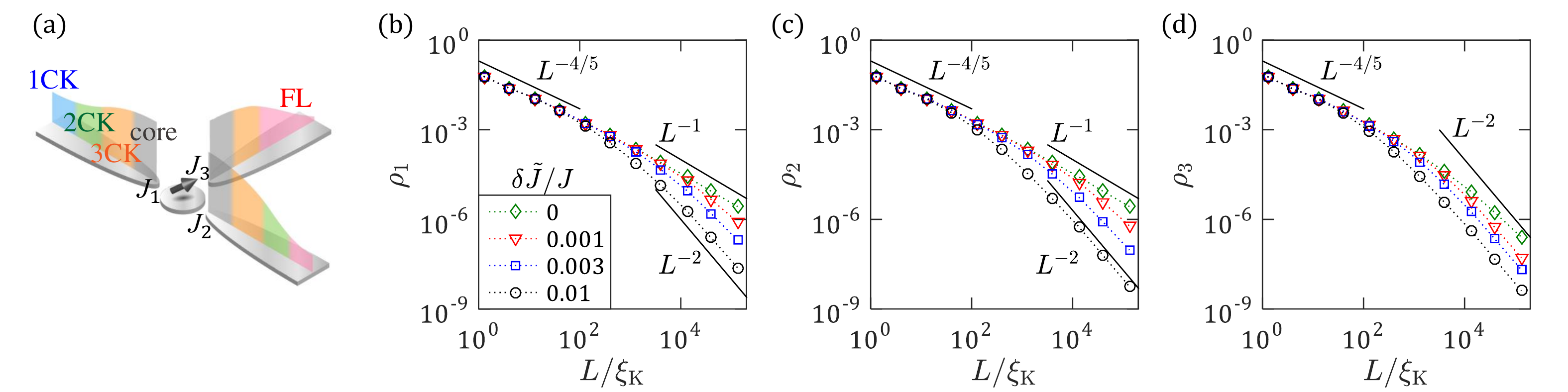}}
\caption{
Shell structure of Kondo clouds in the anisotropic three-channel Kondo (3CK) effect with $J_1 > J_2 > J_3$. (a) Schematic view of the cloud distribution.
(b-d) Distribution  $\rho_1$, $\rho_2$, and $\rho_3$ for different channel anisotropy $\delta \tilde{J}$. Different shells are identified by using the power law exponents of the dependence of the cloud distribution on $L$.
We choose $J_1=J(1+\delta J/2J)(1+\delta \tilde{J}/J)$, $J_2=J(1+\delta J/2J)(1-\delta \tilde{J}/J)$, $J_3=J-\delta J$ where $J=0.3D$, $\delta J=0.01J$, $B=0.1D$, and the $z$-averaging of $z=0$ and $1/2$.
}
\label{MCKC_supp_123_T}
\end{figure*}
 

The result of the channel-anisotropic 3CK effect is well generalized to the shell structure of the Kondo cloud distribution in the $k$CK effects with $k \ge 4$. The core region is followed by the next inner $k$CK shell. Then there occur 
$k''$CK, $q_1$CK, $q_2$CK, $\cdots$, $k$CK shells from the outermost to innermost shells, where $1 \le k'' < q_1 < q_2 < \cdots < k$. 
The values of 
$k'' < q_1 < q_2 < \cdots < k$ are determined by the relative Kondo coupling strengths.
In each intermediate shell, e.g., in the $q$CK shell, the channels $1,2,\cdots,q$ of stronger coupling strengths show $q$CK region, 
while the other channels of weaker coupling exhibit the non-Kondo Fermi liquid region.

\newpage
\begin{center}
\textbf{Supplementary Note 5. DERIVATION OF EQ.~(5)}
\end{center}

We derive Eq.~(5) of the main text. To do this, we first derive the magnetization $\langle \mathbf{S}_{\mathrm{imp}} \rangle$ of the impurity in the presence of the local spin symmetry breaking $H_\textrm{LSB}$.
The $k$CK Hamiltonian $H_{k\mathrm{CK}} = \sum_{j=1}^k  J_j \mathbf{S}_\mathrm{imp} \cdot \mathbf{S}_j(0) + H_{j}$ with isotropic couplings $J_{1} = \cdots = J_{k} = J$ at $T \ll T_{K}$ is described by the boundary conformal field theory (BCFT) Hamiltonian
\begin{equation}\label{BCFT Hamiltonian}
H_{\mathrm{BCFT}} = H_{\mathrm{FP}} + \lambda H_{\mathrm{LI}}.
\end{equation}
$H_{\mathrm{FP}}$ is the fixed-point Hamiltonian invariant under $\mathrm{U}(1) \times \mathrm{SU}(2)_{k} \times \mathrm{SU}(k)_{2}$ Kac-Moody algebra. $\lambda H_{\mathrm{LI}}$ is the leading irrelevant operator with coupling constant $\lambda \propto 1 / T_{\mathrm{K}}^{\Delta}$. According to the Kac-Moody symmetry,  operators are labeled by the representation (quantum numbers of charge $Q$, spin $J_{s}$, flavor $J_{f}$) of three sectors $\mathrm{U}(1)$, $\mathrm{SU}(2)_{k}$, $\mathrm{SU}(k)_{2}$. The identity operator $I$ corresponds to the trivial representation. The spin (resp. flavor)-adjoint primary boundary operator $\pmb{\upphi}_{s}$ (resp. $\pmb{\upphi}_{f}$) corresponds to adjoint representation of $\mathrm{SU}(2)_{k}$ (resp. $\mathrm{SU}(k)_{2}$) with conformal dimension $\Delta = 2 / (k+2)$ (resp. $1- \Delta$).
The local spin symmetry breaking perturbation $H_{\mathrm{LSB}}$ in the $n$-th channel, shown in Eq.~\eqref{LSBxdirection}, has the equivalent form of  
\begin{align}\label{eq:decoh2}
H_\mathrm{LSB} &= \frac{B}{2} \big(\psi^\dagger_{\uparrow n}(L) \psi_{\downarrow n}(
L) + \psi^\dagger_{\downarrow n}(L) \psi_{\uparrow n}(L)\big).
\end{align}
Here $\alpha = \uparrow, \downarrow$ represents the eigenspin states of the spin operator $S^z$ in the $z$ direction. 
The perturbation acts like a local magnetic field at position $x=L$ in the $x$-direction as shown in Eq.~\eqref{LSBxdirection} so that the magnetization of the impurity has only the $x$-direction component $\langle S^x_\textrm{imp} \rangle$; namely
\begin{align}\label{y,z expectation}
\langle S_\mathrm{imp}^{y,z}\rangle=0.
\end{align}
Below we derive $\langle S^x_\textrm{imp} \rangle$, based on BCFT. The result is in good agreement with the NRG calculation in Supplementary Fig.~\ref{MCKC_supp_mag}.

\begin{figure*}[b]
	\centerline{\includegraphics[width=.98\textwidth]{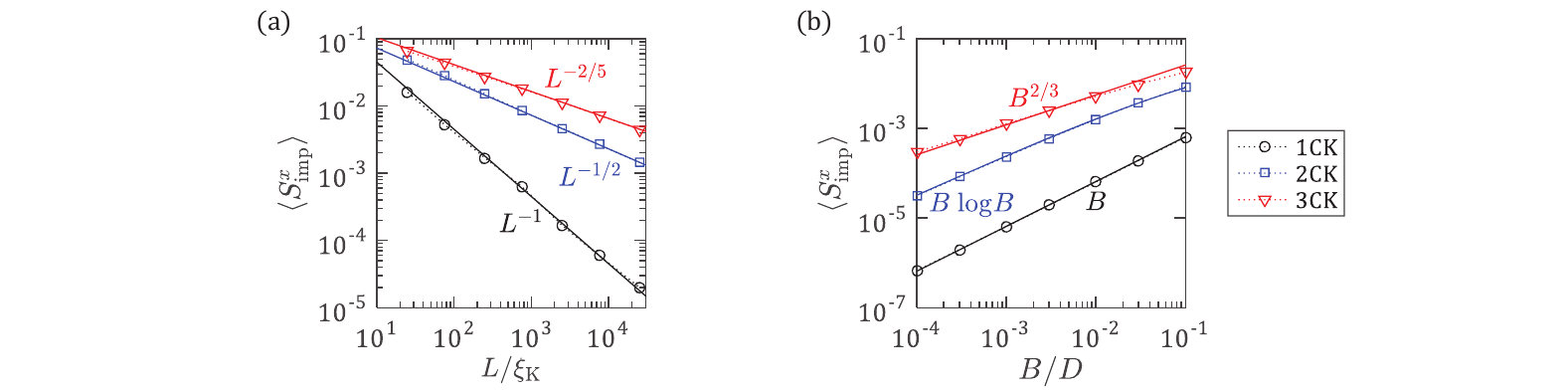}}
	\caption{
	The expectation value $\langle S_\mathrm{imp}^x\rangle$ of the impurity spin in the $x$ direction at zero temperature, in the single-channel Kondo (1CK) (black), the isotropic two-channel Kondo (2CK) (blue), and the isotropic three-channel Kondo (3CK) (red) effect in the presence of the spin symmetry breaking perturbation at position $L$ in the $n=1$ channel.  Its dependence on the perturbation position $L$ and strength $B$ is shown in (a) and (b). 
		Data points, obtained by the numerical renormalization group (NRG) calculation, agree with the fitting curves predicted by the boundary conformal field theory (BCFT).
		In the NRG, we use $J=0.3D$, $B=0.1D$ [for computing the result in (a)], and the $z$-averaging of $z=0$ and $1/2$.
		}
	\label{MCKC_supp_mag}
\end{figure*}

In the derivation, we use a chiral field. We decompose the field $\psi_{\alpha j}(x)$ annihilating an electron with spin $\alpha$ at channel $j$ into the left and right moving chiral fields, $\psi_{\mathrm{L} \alpha j}(x)$ and $\psi_{\mathrm{R} \alpha j}(x)$, 
\begin{equation}\label{Chiral decomposition}
\psi_{\alpha j}(x) = e^{- i k_{F} x} \psi_{\mathrm{L} \alpha j}(x) + e^{ik_{F} x} \psi_{\mathrm{R} \alpha j}(x) \qquad  \alpha = \uparrow,\downarrow, \quad  j = 1, \cdots, k,
\end{equation}
for $x \geq 0$. By defining $\psi_{\mathrm{L} \alpha j}(x) = \psi_{\mathrm{R} \alpha j}(-x)$ for $x < 0$, $\psi_{\alpha j}(x)$ is expressed in terms of one chiral field as $\psi_{\alpha j}(x) = e^{-ik_{F}x} \psi_{\mathrm{L} \alpha j}(x) + e^{ik_{F}x} \psi_{\mathrm{L} \alpha j}(-x)$. 

\newpage
\subsection{A. Multichannel Kondo Model}

We calculate $\langle S_{\mathrm{imp}}^{x} \rangle$ in the case of $k \geq 2$. In this case, the leading irrelevant operator $H_{\mathrm{LI}} = \mathbf{J}_{-1} \cdot \pmb{\upphi}_{s}$ is the Kac-Moody descendant of the spin adjoint primary boundary operator $\pmb{\upphi}_{s}$,  and non-perturbative treatment is required in computing $\langle S_{\mathrm{imp}}^{x} \rangle$ at zero temperature. In the Lagrangian description, the action corresponding to the Hamiltonian $H_{\mathrm{FP}} + \lambda H_{\mathrm{LI}} + H_{\mathrm{LSB}}$ is written as
\begin{equation}\label{action1}
\begin{aligned}
S &= S_{\mathrm{FP}} + \lambda \int_{-\infty}^{\infty} d\tau \, \mathbf{J}_{-1} \cdot \pmb{\upphi}_{s}(\tau,0) \\
&\qquad + \frac{B}{2} \int_{-\infty}^{\infty} d\tau \, \Bigg(J_{\mathrm{L} n}^{x}(\tau,L) + J_{\mathrm{L}n}^{x}(\tau,-L) + e^{2ik_{F}L} \Big(\psi_{\mathrm{L} \uparrow n}^{\dagger}(\tau,L) \psi_{\mathrm{L} \downarrow n}(\tau,-L) + \psi_{\mathrm{L} \downarrow n}^{\dagger}(\tau,L) \psi_{\mathrm{L} \uparrow n}(\tau,-L)\Big) \\
&\qquad \qquad \qquad \qquad \qquad \qquad \qquad \qquad \qquad + e^{- 2ik_{F}L} \Big(\psi_{\mathrm{L} \uparrow n}^{\dagger}(\tau,-L) \psi_{\mathrm{L} \downarrow n}(\tau,L) + \psi_{\mathrm{L} \downarrow n}^{\dagger}(\tau,-L) \psi_{\mathrm{L} \uparrow n}(\tau,L)\Big)\Bigg).
\end{aligned}
\end{equation}
$S_{\mathrm{FP}}$ is the action at the fixed point corresponding to $H_{\mathrm{FP}}$. $\tau$ is the Euclidean time. $J_{\mathrm{L} n}^{x}(\tau,x) \equiv \psi_{\mathrm{L} \uparrow n}^{\dagger}(\tau,x) \psi_{\mathrm{L} \downarrow n}(\tau,x) + \psi_{\mathrm{L} \downarrow n}^{\dagger}(\tau,x) \psi_{\mathrm{L} \uparrow n}(\tau,x)$ is the $x$-component of the local spin. The terms proportional to $B$ come from $H_\textrm{LSB}$, and they are written in terms of the chiral field $\psi_{\alpha j}(x) = e^{-ik_{F}x} \psi_{\mathrm{L} \alpha j}(x) + e^{ik_{F}x} \psi_{\mathrm{L} \alpha j}(-x)$ defined below Eq.~\eqref{Chiral decomposition}. 
We obtain $\langle S_{\mathrm{imp}}^{x} \rangle$ by using the series expansion in the perturbation theory,  
\begin{equation}\label{Series1}
\langle S_{\mathrm{imp}}^{x} \rangle = \frac{1}{\mathcal{Z}} \sum_{m_{1},m_{2},m_{3} = 0}^{\infty} \frac{1}{m_{1}! m_{2}! m_{3}!} A_{m_{1},m_{2},m_{3}},
\end{equation}
\begin{equation}\label{component1}
\begin{aligned}
A_{m_{1},m_{2},m_{3}} &= \Bigg\langle S_{\mathrm{imp}}^{x} \Bigg(- \lambda \int_{-\infty}^{\infty} d\tau \, \mathbf{J}_{-1} \cdot \pmb{\upphi}_{s}(\tau,0)\Bigg)^{m_{1}} \Bigg(- \frac{B}{2} \int_{-\infty}^{\infty} d\tau \, \Big(J_{\mathrm{L}n}^{x}(\tau,L) + J_{\mathrm{L}n}^{x}(\tau,-L)\Big)\Bigg)^{m_{2}} \\
&\qquad \qquad \times \Bigg(- \frac{B}{2} \int_{-\infty}^{\infty} d\tau \, \Big[e^{2ik_{F}L} \Big(\psi_{\mathrm{L} \uparrow n}^{\dagger}(\tau,L) \psi_{\mathrm{L} \downarrow n}(\tau,-L) + \psi_{\mathrm{L} \downarrow n}^{\dagger}(\tau,L) \psi_{\mathrm{L} \uparrow n}(\tau,-L)\Big) \\
&\qquad \qquad \qquad \qquad \qquad \qquad + e^{- 2ik_{F}L} \Big(\psi_{\mathrm{L} \uparrow n}^{\dagger}(\tau,-L) \psi_{\mathrm{L} \downarrow n}(\tau,L) + \psi_{\mathrm{L} \downarrow n}^{\dagger}(\tau,-L) \psi_{\mathrm{L} \uparrow n}(\tau,L)\Big)\Big]\Bigg)^{m_{3}} \Bigg\rangle_{\mathrm{FP}}.
\end{aligned}
\end{equation}
$\langle \cdots \rangle_{\mathrm{FP}}$ is the expectation value with respect to the fixed point action $S_{\mathrm{FP}}$. $\mathcal{Z}$ is the partition function of the action $S$ in Eq.~\eqref{action1}.
The impurity spin is identified with the spin adjoint primary operator $\phi_{s}^{x}$,
\begin{equation}\label{Identification}
S_{\mathrm{imp}}^{x} = \frac{s_{1}}{T_{\mathrm{K}}^{\Delta}} \phi_{s}^{x} + \sum_{i} \psi_{i},
\end{equation}
where $s_{1}$ is a constant and $\psi_{i}$'s indicate the scaling fields of dimension $> \Delta$. 
In the expression of $A_{m_{1},m_{2},m_{3}}$, we rescale the time and space by $(\tau,x) \rightarrow (\tau',x') = (a/L) (\tau,x)$ with $a/L \ll 1$. Using the covariance of the fields under the scale transformation, $A_{m_{1},m_{2},m_{3}}$ becomes
\begin{equation}\label{component2}
\begin{aligned}
A_{m_{1},m_{2},m_{3}} &= \frac{1}{(L/a)^{\Delta + m_{1} \Delta}} \Bigg\langle S_{\mathrm{imp}}^{x} \Bigg(- \lambda \int_{-\infty}^{\infty} d\tau' \, \mathbf{J}_{-1} \cdot \pmb{\upphi}_{s}(\tau',0)\Bigg)^{m_{1}} \Bigg(- \frac{B}{2} \int_{-\infty}^{\infty} d\tau' \, \Big(J_{\mathrm{L}n}^{x}(\tau',a) + J_{\mathrm{L}n}^{x}(\tau',-a)\Big)\Bigg)^{m_{2}} \\
&\qquad \qquad \qquad \qquad \quad \times \Bigg(- \frac{B}{2} \int_{-\infty}^{\infty} d\tau \, \Big[e^{2ik_{F}L} \Big(\psi_{\mathrm{L} \uparrow n}^{\dagger}(\tau',a) \psi_{\mathrm{L} \downarrow n}(\tau',-a) + \psi_{\mathrm{L} \downarrow n}^{\dagger}(\tau',a) \psi_{\mathrm{L} \uparrow n}(\tau',-a)\Big) \\
&\qquad \qquad \qquad \qquad \qquad \qquad + e^{- 2ik_{F}L} \Big(\psi_{\mathrm{L} \uparrow n}^{\dagger}(\tau',-a) \psi_{\mathrm{L} \downarrow n}(\tau',a) + \psi_{\mathrm{L} \downarrow n}^{\dagger}(\tau,-a) \psi_{\mathrm{L} \uparrow n}(\tau',a)\Big)\Big]\Bigg)^{m_{3}} \Bigg\rangle_{\mathrm{FP}}.
\end{aligned}
\end{equation}
Here we utilized the fact that the scaling dimensions of $\phi_{s}^{x}$ and $\mathbf{J}_{-1} \cdot \pmb{\upphi}_{s}$ are $\Delta = 2 / (k+2)$ and $1 + \Delta$, respectively.
In the BCFT description of the multichannel Kondo model~\cite{Ludwig94_supp}, we have the operator product expansion (OPE)
\begin{equation}\label{OPE}
\psi_{\mathrm{L} \uparrow n}^{\dagger}(\tau',a) \psi_{\mathrm{L} \downarrow n}(\tau',-a) + \psi_{\mathrm{L} \downarrow n}^{\dagger}(\tau',a) \psi_{\mathrm{L} \uparrow n}(\tau',-a) \rightarrow  \frac{c_{1}}{a^{1-\Delta}} \phi_{s}^{x}(\tau',0) + \cdots
\end{equation}
where $c_{1}$ is a constant and $\cdots$ represents the scaling fields with dimension larger than $\Delta = 2/(k+2)$. 
Applying this to Eq.~\eqref{component2}, we obtain
\begin{equation}
\begin{aligned}
A_{m_{1},m_{2},m_{3}} &= \frac{1}{(L/a)^{\Delta + m_{1} \Delta}} \Bigg\langle S_{\mathrm{imp}}^{x} \Bigg(- \lambda \int_{-\infty}^{\infty} d\tau' \, \mathbf{J}_{-1} \cdot \pmb{\upphi}_{s}(\tau',0)\Bigg)^{m_{1}} \Bigg(- \frac{B}{2} \int_{-\infty}^{\infty} d\tau' \, \Big(J_{\mathrm{L}n}^{x}(\tau',a) + J_{\mathrm{L}n}^{x}(\tau',-a)\Big)\Bigg)^{m_{2}} \\
&\qquad \qquad \qquad \qquad \quad \times \Bigg(- \frac{B}{2} \int_{-\infty}^{\infty} d\tau \, \Big[e^{2ik_{F}L} \Big(\frac{c_{1}}{a^{1 - \Delta}} \phi_{s}^{x}(\tau',0)\Big) + e^{- 2ik_{F}L} \Big(\frac{c_{1}}{a^{1 - \Delta}} \phi_{s}^{x}(\tau',0)\Big)\Big]\Bigg)^{m_{3}} \Bigg\rangle_{\mathrm{FP}}
\end{aligned}
\end{equation}
We again rescale the time and space by $(\tau',x') \rightarrow (\tau,x) = (L/a) (\tau',x')$ and return to the original ones.
\begin{equation}\label{component3}
\begin{aligned}
A_{m_{1},m_{2},m_{3}} &= \Bigg\langle S_{\mathrm{imp}}^{x} \Bigg(- \lambda \int_{-\infty}^{\infty} d\tau \, \mathbf{J}_{-1} \cdot \pmb{\upphi}_{s}(\tau,0)\Bigg)^{m_{1}} \Bigg(- \frac{B}{2} \int_{-\infty}^{\infty} d\tau \, \Big(J_{\mathrm{L} n}^{x}(\tau,L) + J_{\mathrm{L} n}^{x}(\tau,-L)\Big)\Bigg)^{m_{2}} \\
&\qquad \qquad \qquad \qquad \qquad \qquad \qquad \qquad \qquad \times \Bigg(- \frac{c_{1} B}{L^{1 - \Delta}} \cos(2k_{F}L) \int_{-\infty}^{\infty} d\tau \, \phi_{s}^{x}(\tau,0)\Bigg)^{m_{3}} \Bigg\rangle_{\mathrm{FP}}.
\end{aligned}
\end{equation}
This means that the expectation value is  perturbed by the leading irrelevant operator $\lambda \mathbf{J}_{-1} \cdot \pmb{\upphi}_{s}$, the magnetic field $B (J_{\mathrm{L}n}^{x}(L) + J_{\mathrm{L} n}^{x}(-L))$, and $(c_{1} B / L^{1-\Delta}) \cos(2k_{F}L) \phi_{s}^{x}(0)$.
Since $\phi_{s}^{x}(\tau',0)$ has smaller scaling dimension than $J_{\mathrm{L} n}^{x}(\tau',\pm a)$, the contribution from $\phi_{s}^{x}(\tau',0)$ dominates over  that from $J_{\mathrm{L} n}^{x}(\tau',\pm a)$. Using Eq.~\eqref{Identification}, we have
\begin{equation}\label{magnetic field}
\frac{c_{1} B}{L^{1-\Delta}} \cos(2k_{F}L) \phi_{s}^{x}(0) \simeq \frac{c_{1} B}{L^{1-\Delta}} \cos(2k_{F}L) \frac{T_{\mathrm{K}}^{\Delta}}{s_{1}} S_{\mathrm{imp}}^{x} \equiv B^{x} S_{\mathrm{imp}}^{x}.
\end{equation}
This shows that the local spin symmetry breaking perturbation $H_\textrm{LSB}$ is equivalent, in the calculation of $\langle S_{\mathrm{imp}}^{x} \rangle$,  with the perturbation by the $x$-directional magnetic field $B^x$ applied to the magnetic impurity at low temperature near the fixed point. It is known~\cite{Affleck91_supp,Andrei84_supp} that the magnetic field results in 
\begin{equation}\label{Expectation2}
\langle S_{\mathrm{imp}}^{x} \rangle \propto \begin{cases} B^{x} \log B^{x} &\quad k = 2 \\ (B^{x})^{2/k} &\quad k \geq 3 \end{cases}
\end{equation}
at zero temperature.
Thus, Eq.~\eqref{magnetic field} and Eq.~\eqref{Expectation2} gives the leading order of $\langle S_{\mathrm{imp}}^{x} \rangle$ :
\begin{equation}\label{x expectation}
\langle S_{\mathrm{imp}}^{x} \rangle \propto \begin{cases} (B \log B) (\xi_{\mathrm{K}} / L)^{1/2} \cos (2k_{F}L) &\quad k = 2 \\ B^{2/k} (\xi_{\mathrm{K}}/L)^{2 / (k+2)} (\cos (2k_{F}L))^{2/k} &\quad k \geq 3 \end{cases}
\end{equation}
where $\xi_{\mathrm{K}} = 1 / T_{\mathrm{K}}$.

The LSB breaks the $\mathrm{SU}(2)$ spin symmetry, making the ground state non-degenerate.
So we can use Eq. (3) of the main text, since it is applicable for a pure state such as the non-degenerate ground state. 
By using Eq.~\eqref{y,z expectation}, we have
\begin{align}\label{multiN}
\mathcal{N}(L,T = 0;n) = \sqrt{1 - \frac{4 \mathbf{M}^{2}}{\hslash^{2}}} = \sqrt{1 - \frac{4 \langle S_{\mathrm{imp}}^{x} \rangle^{2}}{\hslash^{2}}} \simeq 1 - \frac{2 \langle S_{\mathrm{imp}}^{x} \rangle^{2}}{\hslash^{2}}, \qquad L \gg \xi_{\mathrm{K}}.
\end{align}
By using Eq.~(2) of the main text and Eqs.~\eqref{x expectation} and \eqref{multiN}, we obtain
\begin{align}\label{multiKCD}
\rho_{n}(L,T = 0) \equiv \mathcal{N}_{0}(T = 0) - \mathcal{N}(L,T = 0;n) \simeq \frac{2 \langle S_{\mathrm{imp}}^{x} \rangle^{2}}{\hslash^{2}} \propto (\frac{\xi_{\mathrm{K}}}{L})^{\frac{4}{k+2}} (\cos (2k_{F}L))^{4/k}, \qquad L \gg \xi_{\mathrm{K}}.
\end{align}
Here we use $\mathcal{N}_{0}(T = 0) = 1$ in the absence of the LSB.
We focus on the envelop of the $L$ dependence in Eq.~\eqref{multiKCD}, because it exhibits a universality of the Kondo cloud:
\begin{align}\label{multiKCD2}
\rho_{n}(L,T = 0) \propto (\frac{\xi_{\mathrm{K}}}{L})^{\frac{4}{k+2}}, \qquad L \gg \xi_{\mathrm{K}} .
\end{align}
It is Eq. (5) for $k \geq 2$ in the main text.
Here we choose the value $2k_{F}L$ as an integer multiple of $\pi$, which give the maximum value $|\cos (2k_{F}L)| = 1$, to trace the envelope and clearly show the power decay of the Kondo cloud.

\subsection{B. Single Channel Kondo Model}

We calculate $\langle S_{\mathrm{imp}}^{x} \rangle$ in the single channel Kondo model. Here, the leading irrelevant operator is given by the local spin product $H_{\mathrm{LI}} = \mathbf{J}_{\mathrm{L}} \cdot \mathbf{J}_{\mathrm{L}}(0)$.  Similar to Eq.~\eqref{action1}, we have the action corresponding to the Hamiltonian $H_{\mathrm{FP}} + \lambda H_{\mathrm{LI}} + H_{\mathrm{LSB}}$ as
\begin{equation}\label{action3}
\begin{aligned}
S &= S_{\mathrm{FP}} + \lambda \int_{-\infty}^{\infty} d\tau \, \mathbf{J}_{\mathrm{L}} \cdot \mathbf{J}_{\mathrm{L}}(\tau,0) \\
&\qquad + \frac{B}{2} \int_{-\infty}^{\infty} d\tau \, \Bigg(J_{\mathrm{L}}^{x}(\tau,L) + J_{\mathrm{L}}^{x}(\tau,-L) + e^{2ik_{F}L} \Big(\psi_{\mathrm{L} \uparrow}^{\dagger}(\tau,L) \psi_{\mathrm{L} \downarrow}(\tau,-L) + \psi_{\mathrm{L} \downarrow}^{\dagger}(\tau,L) \psi_{\mathrm{L} \uparrow}(\tau,-L)\Big) \\
&\qquad \qquad \qquad \qquad \qquad \qquad \qquad \qquad \qquad + e^{- 2ik_{F}L} \Big(\psi_{\mathrm{L} \uparrow}^{\dagger}(\tau,-L) \psi_{\mathrm{L} \downarrow}(\tau,L) + \psi_{\mathrm{L} \downarrow}^{\dagger}(\tau,-L) \psi_{\mathrm{L} \uparrow}(\tau,L)\Big)\Bigg)
\end{aligned}
\end{equation}
Here we omit the channel index $n$ because there is only one channel.  The impurity spin is identified with the spin density operator of scaling dimension 1,
\begin{equation}\label{Identification2}
S_{\mathrm{imp}}^{x} = \frac{s_{2}}{T_{\mathrm{K}}} J_{\mathrm{L}}^{x} + \sum_{i} \psi_{i},
\end{equation}
where $s_{2}$ is a constant and $\psi_{i}$'s indicate the scaling fields of dimension $> 1$. 
In this 1CK case, perturbative treatment with respect to the terms proportional to $B$ is possible in computing $\langle S_{\mathrm{imp}}^{x} \rangle$ at zero temperature, contrary to the $k$CK case with $k \ge 2$.
The leading contribution in the perturbation expansion of $\langle S_{\mathrm{imp}}^{x} \rangle$ with respect to the action terms proportional to $B$ is given by
\begin{equation}\label{x expectation2}
\begin{aligned}
\langle S_{\mathrm{imp}}^{x} \rangle &\simeq - \int_{-\infty}^{\infty} d\tau \, \Bigg\langle S_{\mathrm{imp}}^{x} \frac{B}{2} \Bigg(e^{2ik_{F}L} \Big(\psi_{\mathrm{L} \uparrow}^{\dagger}(\tau,L) \psi_{\mathrm{L} \downarrow}(\tau,-L) + \psi_{\mathrm{L} \downarrow}^{\dagger}(\tau,L) \psi_{\mathrm{L} \uparrow}(\tau,-L)\Big) \\
&\qquad \qquad \qquad \qquad \qquad + e^{- 2ik_{F}L} \Big(\psi_{\mathrm{L} \uparrow}^{\dagger}(\tau,-L) \psi_{\mathrm{L} \downarrow}(\tau,L) + \psi_{\mathrm{L} \downarrow}^{\dagger}(\tau,-L) \psi_{\mathrm{L} \uparrow}(\tau,L)\Big)\Bigg) \Bigg\rangle \\
&\simeq - \int_{-\infty}^{\infty} d\tau \, \Bigg\langle \frac{s_{2}}{T_{\mathrm{K}}} J_{\mathrm{L}}^{x}(0,0) \frac{B}{2} \Bigg(e^{2ik_{F}L} \Big(\psi_{\mathrm{L}\uparrow}^{\dagger}(\tau,L) \psi_{\mathrm{L} \downarrow}(\tau,-L) + \psi_{\mathrm{L} \downarrow}^{\dagger}(\tau,L) \psi_{\mathrm{L} \uparrow}(\tau,-L)\Big) \\
&\qquad \qquad \qquad \qquad \qquad \qquad + e^{- 2ik_{F}L} \Big(\psi_{\mathrm{L} \uparrow}^{\dagger}(\tau,-L) \psi_{\mathrm{L} \downarrow}(\tau,L) + \psi_{\mathrm{L} \downarrow}^{\dagger}(\tau,-L) \psi_{\mathrm{L} \uparrow}(\tau,L)\Big)\Bigg) \Bigg\rangle \\
&\propto \frac{\pi s_{2}}{T_{K}L} B \cos (2k_{F}L).
\end{aligned}
\end{equation}
In the last equality, we used the three-point correlator $\langle J_\textrm{L}^x \psi^\dagger \psi \rangle$ in BCFT.
Hence, $\langle S_{\mathrm{imp}}^{x} \rangle \sim O(B (\xi_{\mathrm{K}} / L))$ in the 1CK.

By following the way to derive Eqs.~\eqref{multiN} and \eqref{multiKCD}, we obtain
\begin{align}
\rho_{n}(L,T = 0) \equiv \mathcal{N}_{0}(T = 0) - \mathcal{N}(L,T = 0;n) \simeq \frac{2 \langle S_{\mathrm{imp}}^{x} \rangle^{2}}{\hslash^{2}} \propto (\frac{\xi_{\mathrm{K}}}{L})^{2} \cos^{2} (2k_{F}L), \qquad L \gg \xi_{\mathrm{K}}.
\end{align}
We focus on the envelope of the Friedel oscillation as in Eqs.~\eqref{multiKCD2} and \eqref{multiKCD}:
\begin{align}
\rho_{n}(L,T = 0) \propto (\frac{\xi_{\mathrm{K}}}{L})^{2}, \qquad L \gg \xi_{\mathrm{K}} .
\end{align}
It is Eq. (5) for the single channel Kondo model in the main text.

\begin{center}
\textbf{Supplementary Note 6. CHANNEL ANISOTROPIC 2CK EFFECTS}
\end{center}
We compute $\langle S_{\mathrm{imp}}^{x} \rangle$ in the channel anisotropic 2CK effect in the presence of the local symmetry breaking perturbation. The total Hamiltonian is  $H + \sum_{n=1,2} H_{\textrm{LSB},n}$. Here $H$ describes the channel anisotropic 2CK effect,
\begin{equation}\label{channel anisotropic Hamiltonian}
H = H_{1} + H_{2} + \sum_{j=1,2} J S_{\mathrm{imp}}^{z} S_{i}^{z}(0) + \sum_{j=1,2} \sum_{\alpha = x,y} (J+(-1)^{i}\delta J) S_{\mathrm{imp}}^{\alpha} S_{j}^{\alpha}(0),
\end{equation}
where  $H_{1(2)}$ is the Hamiltonian for free electrons in the channel 1 (2). 
$H_{\textrm{LSB},n}$ describes the local symmetry breaking perturbation on channel $n=1,2$ in Eq.~\eqref{eq:decoh2}.
Using the chiral field in Eq.~\eqref{Chiral decomposition}, we rewrite $H_{\mathrm{LSB},n}$,
\begin{equation}
\begin{aligned}
H_{\mathrm{LSB},n} &= \frac{B}{2} \Bigg(J_{\mathrm{L}n}^{x}(L) + J_{\mathrm{L}n}^{x}(-L) + e^{2ik_{F}L} \Big(\psi_{\mathrm{L} \uparrow n}^{\dagger}(L) \psi_{\mathrm{L} \downarrow n}(-L) + \psi_{\mathrm{L} \downarrow n}^{\dagger}(L) \psi_{\mathrm{L} \uparrow n}(-L)\Big) \\
&\qquad \qquad \qquad \qquad \qquad + e^{- 2ik_{F}L} \Big(\psi_{\mathrm{L} \uparrow n}^{\dagger}(-L) \psi_{\mathrm{L} \downarrow n}(L) + \psi_{\mathrm{L} \downarrow n}^{\dagger}(-L) \psi_{\mathrm{L} \uparrow n}(L)\Big)\Bigg).
\end{aligned}
\end{equation}

We use the bosonization method~\cite{Emergy92_supp,Zarand00_supp}. The fermion field is bosonized, $\psi_{\mathrm{L} \alpha j}(x) \propto e^{-i \varphi_{\alpha j}(x)}$. The fields $\varphi_{c} = \frac{1}{2} (\varphi_{\uparrow 1} + \varphi_{\downarrow 1} + \varphi_{\uparrow 2} + \varphi_{\downarrow 2})$, $\varphi_{s} = \frac{1}{2} (\varphi_{\uparrow 1} - \varphi_{\downarrow 1} + \varphi_{\uparrow 2} - \varphi_{\downarrow 2})$, $\varphi_{f} = \frac{1}{2} (\varphi_{\uparrow 1} + \varphi_{\downarrow 1} - \varphi_{\uparrow 2} - \varphi_{\downarrow 2})$, $\varphi_{sf} = \frac{1}{2} (\varphi_{\uparrow 1} - \varphi_{\downarrow 1} - \varphi_{\uparrow 2} + \varphi_{\downarrow 2})$ in charge, spin, flavor, spin-flavor sectors are introduced. Emery-Kivelson transformation $U = e^{i S_{\mathrm{imp}}^{z} \varphi_{s}(0)}$ is applied to $H$, $\widetilde{H} = U H U^{\dagger}$. Refermionization decomposes $\widetilde{H}$ into three free fermion Hamiltonians and one resonant level model,
\begin{equation}
\begin{aligned}
UHU^{\dagger} &= \sum_{A = c,s,f,sf} \int dx \, \psi_{A}^{\dagger}(x) i \partial \psi_{A}(x) + \sqrt{\Gamma} (\psi_{sf}(0) + \psi_{sf}^{\dagger}(0))(c_{d} - c_{d}^{\dagger}) + \sqrt{\delta \Gamma} (\psi_{sf}(0) - \psi_{sf}^{\dagger}(0))(c_{d} + c_{d}^{\dagger})
\end{aligned}
\end{equation}
where $c_{d}$ is a local pseudofermion describing the impurity spin through $S_{\mathrm{imp}}^{z} = c_{d}^{\dagger} c_{d} - 1/2$ and $S_{\mathrm{imp}}^{-} = S_{\mathrm{imp}}^{x} - i S_{\mathrm{imp}}^{y} = F_{s} c_{d}$ with the Klein factor $F_s$ of the spin sector, $\psi_{A}(x) \propto e^{-\varphi_{A}(x)}$ is a fermion field in the $A = c,s,f,sf$ sector, $\Gamma = J^{2} / 4a$, $\delta \Gamma = (\delta J)^{2} / 4a$, and $a$ is a short distance cutoff.
 The impurity spin operator and the fermion field are decomposed into Majorana fermions $\hat{a}$, $\hat{b}$, $\chi_{A}$, $\eta_{A}$ as
\begin{equation}\label{impurity Majorana} 
\hat{a} = \frac{c_{d} + c_{d}^{\dagger}}{\sqrt{2}}, \qquad \hat{b} = \frac{c_{d} - c_{d}^{\dagger}}{\sqrt{2} i}, \qquad \chi_{A}(x) = \frac{\psi_{A}(x) + \psi_{A}^{\dagger}(x)}{\sqrt{2}}, \qquad \eta_{A}(x) = \frac{\psi_{A}(x) - \psi_{A}^{\dagger}(x)}{\sqrt{2} i}, \qquad A = c,s,f,sf.
\end{equation}
The Hamiltonian is rewritten in terms of the Majorana fermions as
\begin{equation}
\begin{aligned}
UHU^{\dagger} &= \sum_{A = c,s,f,sf} \Bigg(\frac{1}{2} \int dx \, \chi_{A}^{\dagger}(x) i \partial \chi_{A}(x) + \frac{1}{2} \int dx \, \eta_{A}^{\dagger}(x) i \partial \eta_{A}(x)\Bigg) + 2 i \sqrt{\Gamma} \chi_{sf}(0) \hat{b} + 2 i \sqrt{\delta \Gamma} \eta_{sf}(0) \hat{a}.
\end{aligned}
\end{equation}
The quadratic Hamiltonian for the spin-flavor sector is diagonalized as
\begin{equation}
\frac{1}{2} \int dx \, \widetilde{\chi}_{sf}^{\dagger}(x) i \partial \widetilde{\chi}_{sf}(x) + \frac{1}{2} \int dx \, \widetilde{\eta}_{sf}^{\dagger}(x) i \partial \widetilde{\eta}_{sf}(x),
\end{equation}
where $\widetilde{\chi}_{sf}$ is related with $\chi_{sf}$ and $\hat{b}$ and $\widetilde{\eta}_{sf}$ is related with $\eta_{sf}$ and $\hat{a}$.
In the limit of low energy ($\ll T_{\mathrm{K}}$) or $\Gamma, \delta \Gamma \to \infty$, the modified Majorana fermions $\widetilde{\chi}_{sf}$ and $\widetilde{\eta}_{sf}$ absorb the Majorana fermions $\hat{b}$ and $\hat{a}$ of the impurity, respectively, satisfying
$\widetilde{\chi}_{sf}(x) = \chi_{sf}(x) \mathrm{sgn}(x)$ and $\widetilde{\eta}_{sf}(x) = \eta_{sf}(x) \mathrm{sgn}(x)$ at $x \ne 0$ and
\begin{equation}\label{absorption}
\widetilde{\chi}_{sf}(0) =  \sqrt{\Gamma}   \hat{b}, \qquad  \widetilde{\eta}_{sf}(0)=  \sqrt{\delta \Gamma} \hat{a}.
\end{equation}
Hence the modified Majorana fermions obey  a modified boundary condition~\cite{Sela09_supp}. 
 After the absorption, the boson field $\widetilde{\varphi}_{sf}(x)$ corresponding to the modified fermion field $\widetilde{\psi}_{sf}(x) \equiv \widetilde{\chi}_{sf}(x) + i \widetilde{\eta}_{sf}(x) \sim F_{sf} e^{-i \widetilde{\varphi}_{sf}(x)}$ satisfies~\cite{Maldacena97_supp,Ye98_supp}
\begin{equation}\label{sf boundary condition}
\widetilde{\varphi}_{sf}(0^+) = \widetilde{\varphi}_{sf}(0^-) + \pi.
\end{equation}
Furthermore, the Emery-Kivelson transformation $U$ is a boundary condition changing operator, and the boson field $\varphi_{s}$ in the spin sector is affected by $U$. The transformed boson field $\widetilde{\varphi}_{s}$ in the spin sector satisfies~\cite{Maldacena97_supp,Ye98_supp}
\begin{equation}\label{s boundary condition}
\widetilde{\varphi}_{s}(0^+) = \widetilde{\varphi}_{s}(0^-) + \pi.
\end{equation}
Therefore, the Hamiltonian $H$ is equivalent with the free theory described by the boson fields $\varphi_{c}$, $\widetilde{\varphi}_{s}$, $\varphi_{f}$ and $\widetilde{\varphi}_{sf}$ with the modified boundary conditions in Eq.~\eqref{sf boundary condition} and Eq.~\eqref{s boundary condition}. 

According to Eqs.~\eqref{impurity Majorana} and \eqref{absorption}, the $x$-component of the impurity spin operator is expressed as
\begin{equation}
\begin{aligned} \label{bosonization impurity spin}
\widetilde{S}_{\mathrm{imp}}^{x} \equiv U S_{\mathrm{imp}}^{x} U^{-1}  &  = e^{i \varphi_{s}(0)} \frac{\hat{a} - i \hat{b}}{2 \sqrt{2}} F_{s}^{\dagger} + F_{s} \frac{\hat{a} + i \hat{b}}{2 \sqrt{2}} e^{-i \varphi_{s}(0)} \\
&= - \frac{1}{4 \sqrt{a}} \Bigg(\frac{1}{\sqrt{\delta \Gamma}} + \frac{1}{\sqrt{\Gamma}}\Bigg) F_{sf} F_{s}^{\dagger} e^{i\widetilde{\varphi}_{s}(0)} e^{-i\widetilde{\varphi}_{sf}(0)} + \frac{1}{4\sqrt{a}} \Bigg(\frac{1}{\sqrt{\delta \Gamma}} - \frac{1}{\sqrt{\Gamma}}\Bigg) F_{sf}^{\dagger} F_{s}^{\dagger} e^{i \widetilde{\varphi}_{s}(0)} e^{i \widetilde{\varphi}_{sf}(0)} \\
&\qquad + \frac{1}{4\sqrt{a}} \Bigg(\frac{1}{\sqrt{\delta \Gamma}} - \frac{1}{\sqrt{\Gamma}}\Bigg) F_{s} F_{sf} e^{- i \widetilde{\varphi}_{sf}(0)} e^{- i \widetilde{\varphi}_{s}(0)} - \frac{1}{4 \sqrt{a}} \Bigg(\frac{1}{\sqrt{\delta \Gamma}} + \frac{1}{\sqrt{\Gamma}}\Bigg) F_{s} F_{sf}^{\dagger} e^{i\widetilde{\varphi}_{sf}(0)} e^{- i\widetilde{\varphi}_{s}(0)}. 
\end{aligned}
\end{equation}
The local spin symmetry breaking perturbation in channel $n$ is Emery-Kivelson transformed, $\widetilde{H}_{\mathrm{LSB},n} = U H_{\mathrm{LSB},n} U^{-1}$,
\begin{equation}\label{transformed LSB 1}
\begin{aligned}
\widetilde{H}_{\mathrm{LSB},1} &= \frac{B}{2} \Bigg[\frac{F_{sf}^{\dagger} F_{s}^{\dagger}}{a} e^{i \widetilde{\varphi}_{s}(L)} e^{i \widetilde{\varphi}_{sf}(L)} + \frac{F_{s} F_{sf}}{a} e^{-i \widetilde{\varphi}_{s}(L)} e^{-i \widetilde{\varphi}_{sf}(L)} \\
&\qquad + \frac{F_{sf}^{\dagger} F_{s}^{\dagger}}{a} e^{i \widetilde{\varphi}_{s}(-L)} e^{i \widetilde{\varphi}_{sf}(-L)} + \frac{F_{s} F_{sf}}{a} e^{-i \widetilde{\varphi}_{s}(-L)} e^{-i \widetilde{\varphi}_{sf}(-L)} \\
&\qquad + e^{2ik_{F}L} \Bigg(\frac{F_{sf}^{\dagger} F_{s}^{\dagger}}{a} e^{\frac{i}{2} (\varphi_{c}(L) + \widetilde{\varphi}_{s}(L) + \varphi_{f}(L) + \widetilde{\varphi}_{sf}(L))} e^{-\frac{i}{2} (\varphi_{c}(-L) - \widetilde{\varphi}_{s}(-L) + \varphi_{f}(-L) - \widetilde{\varphi}_{sf}(-L))} \\
&\qquad \qquad \qquad \qquad + \frac{F_{s} F_{sf}}{a} e^{\frac{i}{2} (\varphi_{c}(L) - \widetilde{\varphi}_{s}(L) + \varphi_{f}(L) - \widetilde{\varphi}_{sf}(L))} e^{- \frac{i}{2} (\varphi_{c}(-L) + \widetilde{\varphi}_{s}(-L) + \varphi_{f}(-L) + \widetilde{\varphi}_{sf}(-L))}\Bigg) \\
&\qquad + e^{-2ik_{F}L} \Bigg(\frac{F_{sf}^{\dagger} F_{s}^{\dagger}}{a} e^{\frac{i}{2} (\varphi_{c}(-L) + \widetilde{\varphi}_{s}(-L) + \varphi_{f}(-L) + \widetilde{\varphi}_{sf}(-L))} e^{-\frac{i}{2} (\varphi_{c}(L) - \widetilde{\varphi}_{s}(L) + \varphi_{f}(L) - \widetilde{\varphi}_{sf}(L))} \\
&\qquad \qquad \qquad \qquad + \frac{F_{s}F_{sf}}{a} e^{\frac{i}{2} (\varphi_{c}(-L) - \widetilde{\varphi}_{s}(-L) + \varphi_{f}(-L) - \widetilde{\varphi}_{sf}(-L))} e^{-\frac{i}{2} (\varphi_{c}(L) + \widetilde{\varphi}_{s}(L) + \varphi_{f}(L) + \widetilde{\varphi}_{sf}(L))}\Bigg)\Bigg],
\end{aligned}
\end{equation}
\begin{equation}\label{transformed LSB 2}
\begin{aligned}
\widetilde{H}_{\mathrm{LSB},2} &= \frac{B}{2} \Bigg[\frac{F_{sf} F_{s}^{\dagger}}{a} e^{i \widetilde{\varphi}_{s}(L)} e^{- i \widetilde{\varphi}_{sf}(L)} + \frac{F_{s} F_{sf}^{\dagger}}{a} e^{-i \widetilde{\varphi}_{s}(L)} e^{i \widetilde{\varphi}_{sf}(L)} \\
&\qquad + \frac{F_{sf} F_{s}^{\dagger}}{a} e^{i \widetilde{\varphi}_{s}(-L)} e^{-i \widetilde{\varphi}_{sf}(-L)} + \frac{F_{s} F_{sf}^{\dagger}}{a} e^{-i \widetilde{\varphi}_{s}(-L)} e^{i \widetilde{\varphi}_{sf}(-L)} \\
&\qquad + e^{2ik_{F}L} \Bigg(- \frac{F_{sf} F_{s}^{\dagger}}{a} e^{\frac{i}{2} (\varphi_{c}(L) + \widetilde{\varphi}_{s}(L) - \varphi_{f}(L) - \widetilde{\varphi}_{sf}(L))} e^{-\frac{i}{2} (\varphi_{c}(-L) - \widetilde{\varphi}_{s}(-L) - \varphi_{f}(-L) + \widetilde{\varphi}_{sf}(-L))} \\
&\qquad \qquad \qquad \qquad - \frac{F_{s} F_{sf}^{\dagger}}{a} e^{\frac{i}{2} (\varphi_{c}(L) - \widetilde{\varphi}_{s}(L) - \varphi_{f}(L) + \widetilde{\varphi}_{sf}(L))} e^{- \frac{i}{2} (\varphi_{c}(-L) + \widetilde{\varphi}_{s}(-L) - \varphi_{f}(-L) - \widetilde{\varphi}_{sf}(-L))}\Bigg) \\
&\qquad + e^{-2ik_{F}L} \Bigg(- \frac{F_{sf} F_{s}^{\dagger}}{a} e^{\frac{i}{2} (\varphi_{c}(-L) + \widetilde{\varphi}_{s}(-L) - \varphi_{f}(-L) - \widetilde{\varphi}_{sf}(-L))} e^{-\frac{i}{2} (\varphi_{c}(L) - \widetilde{\varphi}_{s}(L) - \varphi_{f}(L) + \widetilde{\varphi}_{sf}(L))} \\
&\qquad \qquad \qquad \qquad - \frac{F_{s} F_{sf}^{\dagger}}{a} e^{\frac{i}{2} (\varphi_{c}(-L) - \widetilde{\varphi}_{s}(-L) - \varphi_{f}(-L) + \widetilde{\varphi}_{sf}(-L))} e^{-\frac{i}{2} (\varphi_{c}(L) + \widetilde{\varphi}_{s}(L) - \varphi_{f}(L) - \widetilde{\varphi}_{sf}(L))}\Bigg)\Bigg].
\end{aligned}
\end{equation}

We calculate $\langle S_{\mathrm{imp}}^{x} \rangle$ using Eqs.~\eqref{bosonization impurity spin}-\eqref{transformed LSB 2}. When the local spin symmetry breaking perturbation is applied only to channel 1,  only the terms proportional to  $e^{2ik_{F}L}$ and $e^{-2ik_{F}L}$ in Eq.~\eqref{transformed LSB 1} contribute in the first order perturbation,  
\begin{equation}\label{local magnetization by LSB 1}
\langle S_{\mathrm{imp}}^{x} \rangle = c \times \frac{B}{L} \Bigg(\frac{1}{\sqrt{\delta \Gamma}} - \frac{1}{\sqrt{\Gamma}}\Bigg) + O(B^{2}),
\end{equation}
where $c$ is a constant. When the local spin symmetry breaking perturbation is applied only to channel 2, we find
\begin{equation}\label{local magnetization by LSB 2}
\langle S_{\mathrm{imp}}^{x} \rangle = c \times \frac{B}{L} \Bigg(\frac{1}{\sqrt{\delta \Gamma}} + \frac{1}{\sqrt{\Gamma}}\Bigg) + O(B^{2}),
\end{equation}
with the same constant $c$. Let $\mathcal{N}_{\mathrm{I} | \mathrm{E}}^{(1)}$ (resp. $\mathcal{N}_{\mathrm{I} | \mathrm{E}}^{(2)}$) be the entanglement negativity between the impurity and the environment when the perturbation is applied to channel 1 (resp. 2). From Eqs.~\eqref{local magnetization by LSB 1} and \eqref{local magnetization by LSB 2}, we obtain
\begin{equation}
\frac{1 - \mathcal{N}_{\mathrm{I} | \mathrm{E}}^{(1)}}{1 - \mathcal{N}_{\mathrm{I} | \mathrm{E}}^{(2)}} = \frac{1 - \sqrt{1 - 4[c \frac{B}{L} (\frac{1}{\sqrt{\delta \Gamma}} - \frac{1}{\sqrt{\Gamma}})]^{2}}}{1 - \sqrt{1 - 4[c \frac{B}{L} (\frac{1}{\sqrt{\delta \Gamma}} + \frac{1}{\sqrt{\Gamma}})]^{2}}} \simeq \Bigg(\frac{\sqrt{\Gamma} - \sqrt{\delta \Gamma}}{\sqrt{\Gamma} + \sqrt{\delta \Gamma}}\Bigg)^{2}.
\end{equation}

\newpage
\begin{center}
\textbf{Supplementary Note 7. EXPERIMENTAL SETUP}
\end{center}

In the main text, we propose to apply a quantum point contact (QPC) on an edge channel at distance $L$ from the metallic dot in the charge Kondo circuit~\cite{Iftikhar15_supp,Iftikhar18_supp} in Fig.~\ref{MCKC_fig4}a.
Below, we show that the QPC causes an LSB, 
and suggest experimental parameters for detection of the entanglement shells of Kondo clouds.

We first discuss why the QPC causes an LSB.
In the circuit, the excess charge of the metallic dot supports an impurity pseudospin 1/2, and electron tunneling between an edge channel and the dot results in pseudospin flip; electron tunneling from the channel to the dot leads to pseudospin flip, saying, from down to up, while tunneling from the dot to the channel results in pseudospin flip from up to down.
The QPC causes scattering of an electron on an edge channel, giving rise to asymmetry between down-to-up and up-to-down pseudospin flips.
It is because the QPC prevents the electron from tunneling to the dot
with the probability determined by the QPC scattering amplitude, reducing the down-to-up pseudospin flip.


We rigorously show this by computing the LDOS of the edge channel at the point where electron tunneling to the dot happens. 
The Hamiltonian of the edge channel is written as $H_\mathrm{edge}+H_\mathrm{QPC}$, where 
$H_\mathrm{edge} = i \int_{-\infty}^{\infty} \psi^{\dagger}(x) \partial \psi(x) \, dx$ describes the edge channel and
$H_\mathrm{QPC}$ describes the QPC,
\begin{equation}\label{QPC}
H_\mathrm{QPC} = B \psi^{\dagger}(L) \psi(-L) + B \psi^{\dagger}(-L) \psi(L) .
\end{equation}
The field operator $\psi(x)$ annihilates an electron at coordinate $x$ in the chiral edge channel,  
$B$ is the electron tunneling strength at the QPC, 
and the tunneling happens between the positions $x=-L$ and $x=L$ on the channel. The Hamiltonian $H_\mathrm{edge}+H_\mathrm{QPC}$ can be diagonalized as $\sum_{\epsilon} \epsilon c_{\epsilon}^{\dagger} c_{\epsilon}$ with $c_{\epsilon}^{\dagger} = \int_{-\infty}^{\infty} f(x) \psi^{\dagger}(x) \, dx$. The commutation relation $[H_\mathrm{edge}+H_\mathrm{QPC},c_{\epsilon}^{\dagger}] = \epsilon c_{\epsilon}^{\dagger}$ yields the equation of motion
\begin{equation}
i \partial f(x) + B f(-L) \delta(x-L) + B f(L) \delta(x+L) = \epsilon f(x) ,
\end{equation}
and the mode matching method gives
\begin{equation}
f(x) \propto \begin{cases} \: \mathrm{exp}(- i \epsilon x) &\quad - \infty < x < -L
\\
\bigg(\dfrac{1 - B^{2}/4}{1 - iB \mathrm{exp}(-2i \epsilon L) + B^{2}/4}\bigg) \: \mathrm{exp}(-i \epsilon x) &\quad - L < x < L
\\
\bigg(\dfrac{1 + iB \mathrm{exp}(2 i \epsilon L) + B^{2}/4}{1 - iB \mathrm{exp}(-2 i \epsilon L) + B^{2}/4}\bigg) \: \mathrm{exp}(- i \epsilon x) &\quad L < x < \infty \end{cases}.
\end{equation}
The LDOS $\nu(\epsilon)$ of the channel at $x = 0$ (the location at which electron tunneling to the metallic dot happens) is found,
\begin{equation}\label{LDOScharge}
\nu(\epsilon) = |f(0)|^{2} = \nu_{0} \frac{(1 - B^{2}/4)^{2}}{(1 - B \sin (2 \epsilon L) + B^{2}/4)^{2} + B^{2} \cos^{2}(2\epsilon L)} = \nu_{0} \Big(1 + 2B \sin(2\epsilon L) + O(B^{2})\Big),
\end{equation}
where $\nu_{0}$ is a constant.
The LDOS $\nu(\epsilon)$ of the edge channel has the same form with the LDOS in Eq.~(\ref{LDOS}).
This shows that the QPC acts as an LSB.

As discussed in the main text,
the excess charges of  the metallic dot 
is equivalent  with the magnetization of a spin Kondo impurity, and the excess charge can be measured by using a charge detector placed near the dot~\cite{Field93_supp}.
The dependence of the excess charge $\Delta Q$ on the distance $L$ 
provides the information of the spatial distribution of the Kondo cloud,
$\rho_n=1-\sqrt{1-4\langle\Delta Q/e\rangle^2} \simeq 2\langle\Delta Q / e \rangle^2$.
We compute the excess charge $\Delta Q$ in the Fig.~\ref{MCKC_fig4}b by solving the Hamiltonian $H^\mathrm{charge}_{k\mathrm{CK}} + H_\mathrm{QPC}$ with the NRG.
$H^\mathrm{charge}_{k\mathrm{CK}}$ describes the $k$CK effect in a charge Kondo circuit \cite{Iftikhar18_supp,Mitchell16_supp}
\begin{equation}
H^\mathrm{charge}_{k\mathrm{CK}} = \sum_{j=1}^k H_j + \sum_{j=1}^k  J_j (S^+_\mathrm{imp} S^-_j + S^-_\mathrm{imp} S^+_j).
\end{equation}
$H_j$ describes the $j$-th chiral edge channel, and $H_\mathrm{QPC}$ is introduced in Eq.~\eqref{QPC}. 
This Hamiltonian can be solved in the NRG approach with applying the LDOS in Eq.~\eqref{LDOScharge}.
The excess charge  $\langle\Delta Q/e\rangle$  is obtained from the magnetization $\langle S^z_\mathrm{imp}\rangle$.
The parameters of the NRG calculation are given in Sec.~I.
The QPC strength $B$ is experimentally obtained from the reflection probability $R$ of the QPC as they are related as $R=B^2/(1+B^2/4)^2$. We set the QPC strength $B$ such that $R = 0.15$ in our computation.

We suggest experimental parameters of a charge Kondo circuit for detection of the entanglement shells of Kondo clouds.
In the realization~\cite{Iftikhar15_supp,Iftikhar18_supp} of a charge Kondo circuit,  the charging energy is about 25 $\mathrm{\upmu eV}$, the Kondo temperature can be tuned over the range of $0.01\, \mathrm{K} \lesssim T_\mathrm{K}\lesssim 10 \, \mathrm{K}$, and the temperature is about 15 mK. This implies that the Kondo cloud length $\xi_\mathrm{K}$ is within the range of $0.08 \, \mathrm{\upmu m} \lesssim \xi_\mathrm{K} \lesssim 80 \, \mathrm{\upmu m}$. Here, the Fermi velocity is assumed as  $v_\mathrm{F} \sim 10^5 \, \mathrm{m/s}$ that is typical in integer quantum Hall edge channels. We propose QPC positions $L$ 
in a range of $3 \, \mathrm{\upmu m} \lesssim L \lesssim 20 \, \mathrm{\upmu m}$, which needs to be shorter than the phase coherence length. Variation of $L$ within the range, combined with variation of $T_\textrm{K}$ over the tuning range, allows one to have $0.03 \lesssim L / \xi_\mathrm{K}  \lesssim 200$, with which one can measure the dependence of the excess charge on  $L / \xi_\mathrm{K}$ in Fig. 4 of the main text, hence, the spatial distribution of Kondo clouds in the isotropic $k$CK effects with $k=1,2,3$. Similarly, in the charge Kondo circuit $T^*$ can be tuned over the range of 
$10 \, \mathrm{mK} \lesssim T^* \lesssim 1 \, \mathrm{K}$. 
It is hence possible to have $\xi^* / \xi_\textrm{K} = T_\textrm{K} / T^* \gtrsim 100$.
This implies that one can detect the shell structure of Kondo clouds in the anisotropic $k$CK effects (with $k=1,2,3$) by tuning $L / \xi^*$ and $L / \xi_\textrm{K}$ over a sufficient range. 
Note that this strategy of varying $L/\xi_\mathrm{K}$ was used in a previous experimental report~\cite{Borzenets20_supp} on observation of a Kondo cloud of the single channel Kondo effect.
Sensing the excess charge of $\Delta Q \lesssim 0.01e$, which is necessary to observe crossovers between different entanglement shells of the cloud, is experimentally feasible~\cite{Pierre_supp}. Here $e$ is the electron charge.
It would be possible to experimentally measure the power-law decay in Eq.~\eqref{powerlaw}, although the power-law behavior is accompanied by the Friedel oscillation, by fine-tuning $2k_{F}L$ as in an experiment~\cite{Borzenets20_supp}.

\vspace{0.5cm}
\begin{center}
\textbf{Supplementary References}
\end{center}

\end{document}